\documentclass[11pt]{article}
\pdfoutput=1

\usepackage{epsfig}
\usepackage{graphicx}
\usepackage[tight]{subfigure}
\usepackage{epstopdf}
\usepackage{amsfonts}
\usepackage{amssymb}
\usepackage{amsthm}
\usepackage{amsmath}
\usepackage{multirow}
\usepackage{array}

\newcommand{\newc}{\newcommand}
\newc{\ra}{\rightarrow}
\newc{\lra}{\leftrightarrow}
\newc{\be}{\begin{equation}}
\newc{\ee}{\end{equation}}
\newc{\bs}{\begin{split}}
\newc{\es}{\end{split}}
\newc{\ba}{\begin{eqnarray}}
\newc{\ea}{\end{eqnarray}}
\newc{\ov}{\overline}
\newc{\pa}{\partial}
\newc{\D}{\Delta}

\newc{\nn}{\nonumber}

\newc{\tref}[1]{Table \ref{#1}}
\newc{\eref}[1]{Equation \eqref{#1}}
\newc{\su}[1]{$SU(#1)$}
\newc{\bm}[1]{\mathbf{#1}}
\newc{\fref}[1]{Figure \ref{#1}}
\begin{document}

\begin{titlepage}
	
	\vspace*{0.7cm}

	\begin{center}
		{
			\bf\LARGE
			R-Parity violation in F-Theory }
		\\[12mm]
		Miguel~Crispim~Rom\~ao,$^{\star}$
		\footnote{E-mail: \texttt{m.crispim-romao@soton.ac.uk}}
		Athanasios~Karozas$^{\dagger}$
		\footnote{E-mail: \texttt{akarozas@cc.uoi.gr}},
		Stephen~F.~King$^{\star}$
		\footnote{E-mail: \texttt{king@soton.ac.uk}},
		George~K.~Leontaris$^{\dagger}$
		\footnote{E-mail: \texttt{leonta@uoi.gr}},
		Andrew~K.~Meadowcroft$^{\star}$
		\footnote{E-mail: \texttt{a.meadowcroft@soton.ac.uk}}
		\\[-2mm]
		
	\end{center}
	\vspace*{0.50cm}
	\centerline{$^{\star}$ \it
		Physics and Astronomy, University of Southampton,}
	\centerline{\it
		SO17 1BJ Southampton, United Kingdom }
	\vspace*{0.2cm}
	\centerline{$^{\dagger}$ \it
		Physics Department, Theory Division, Ioannina University,}
	\centerline{\it
		GR-45110 Ioannina, Greece}
	\vspace*{1.20cm}
	
	\begin{abstract}
		\noindent
		We discuss R-parity violation (RPV) in semi-local and local F-theory constructions.
We first present a detailed analysis of all possible combinations
of RPV operators arising from semi-local F-theory spectral cover constructions, assuming an $SU(5)$ GUT. We provide a classification of all possible allowed 
combinations of RPV operators originating from operators of the form
$10\cdot \bar 5\cdot \bar 5$,
including the effect of $U(1)$ fluxes with global restrictions.
We then relax the global constraints and 
perform explicit computations of the bottom/tau and RPV 
Yukawa couplings, 
at an $SO(12)$ local point of enhancement in the presence of 
general fluxes subject only to local flux restrictions.
We compare our results to the experimental limits on each allowed RPV operator,
and show that operators such as $LLe^c$, $LQd^c$ and $u^cd^cd^c$ may be present 
separately within current bounds, possibly on the edge of observability, suggesting 
lepton number violation or neutron-antineutron oscillations
could constrain F-theory models.

	\end{abstract}
	
\end{titlepage}

\tableofcontents

\section{Introduction}

The quest for a unified theory of elementary particles has led to numerous extensions 
of the successful Standard Model (SM) of electroweak and strong interactions. During the last decades, 
string theory has been proven to be a powerful approach to describing gravity, which also
enforces restrictions on the particle physics theory. Grand Unified Theories (GUTs)~\cite{Georgi:1974sy} may be embedded
in string scenarios, while supersymmetry (SUSY) is also incorporated in a consistent way,
leading to a natural solution of the hierarchy problem.
Although string theory does not provide a unique prediction
for the precise GUT symmetry and matter content, it enables 
a classification of possible solutions in a well defined and organised way.
Moreover, it provides computational tools
for various parameters such as the Yukawa couplings and potentials which would otherwise be 
left unspecified in more arbitrary extensions of the Standard Model.

Among other restrictions imposed by string theory principles,
of particular importance are those  on the massless spectrum. 
In many string constructions only small representations 
such as the fundamental and spinorial of the GUT group are available 
while the adjoint or higher ones are absent in the massless spectrum.
In some cases this puts model building in a precarious position
since the spontaneous breaking of most successful GUTs requires Higgs fields
in the adjoint representation. But it was precisely this difficulty which 
gave rise to the invention of new symmetry breaking mechanisms and other alternative ways
to obtain the Standard Model.
In the case of $SU(5)$ for example~\cite{Georgi:1974sy}, one manages  to circumvent this obstacle
by replacing it with the flipped -$SU(5)\times U(1)$- version 
 of the model~\cite{Barr:1981qv},\cite{Antoniadis:1987dx}, while in the 
case of Pati-Salam symmetry $SU(4)\times SU(2)\times SU(2)$~\cite{Pati:1974yy}  the  adjoint Higgs field,
which transforms under the gauge group as 
$(15,1,1)$, is replaced by the vector-like Higgs pair of fields
which transform as 
$(4,1,2)+(\bar 4,1,2)$~\cite{Antoniadis:1988cm},\cite{King:1997ia}. 
Analogously, a way out of this difficulty in F-theory models~\cite{Donagi:2008ca, Donagi:2008kj, Beasley:2008dc, Beasley:2008kw}, where the singularity is
realised on a del Pezzo surface, is the use of fluxes to break
the GUT symmetry. 
Indeed, in the last decade or so, a considerable amount of work has been devoted to
the possibility of successfully embedding GUTs such as $SU(5)$ as well as exceptional $E_{6,7,8}$
in an F-theory
framework, leading to new features
\cite{Heckman:2010bq, Weigand:2010wm, Leontaris:2012mh, Maharana:2012tu}. 

Recently, some of us have analysed various phenomenological aspects of F-theory effective models
using the spectral cover description~\cite{Karozas:2014aha,Karozas:2015zza,Romao:2015jrh}.
While, in F-constructions, R-parity conservation (RPC) can emerge 
either as a remnant symmetry of extra $U(1)$ factors, or it can be imposed 
by appealing to some geometric property of the internal manifold and the flux \cite{Hayashi:2009bt},
there is no compelling reason to assume this. 
Moreover, experimental bounds permit R-parity violating (RPV) 
interactions at small but non-negligible rates, providing a generic signature of F-theory models.
In the field theory context, 
RPV proved to be the {\it Achilles heel} of many SUSY GUTs. 
The most dangerous such couplings induce the tree-level operators
$QLd^c, d^cd^cu^c, e^cLL$ and in the absence of a suitable
symmetry or displacement mechanism, all of them appearing simultaneously
can lead to Baryon and Lepton (B and L) violating processes at unacceptable rates~\cite{Dimopoulos:1988jw}. 
On the other hand, in F-theory constructions, parts of GUT multiplets are typically projected out by fluxes, 
giving rise only to a part of the above operators. In other cases, due to symmetry arguments, the Yukawa couplings relevant to RPV operators are identically zero. 
As a result, several B/L violating processes, either are completely prevented or occur at lower rates
in F-theory models, providing a controllable signal of RPV. This observation motivates a general study 
of RPV in F-theory, which is the subject of this paper.

In the present paper, then, we consider RPV in local F-theory, trying to be as general  as possible, with the goal of making a bridge between F-theory and experiment. 
An important goal of the paper is to compute the strength of the RPV Yukawas couplings, which mainly depend on 
the topological properties of the internal space and are more or less independent of 
many details of a particular model, enabling us to work in a generic local F-theory setting.
We focus on F-theory SU(5) constructions, where a displacement mechanism, 
based on non-trivial fluxes, renders several GUT multiplets incomplete. 
This mechanism has already been suggested to eliminate the colour triplets from the Higgs five-plets, 
so that dangerous dimension-5 proton decay operators are not present. However, it turns out that, 
in several cases, not only the Higgs but also other matter multiplets are incomplete, while 
the superpotential structure  is such that it implies RPV terms. 
In this context, it is quite common that not all of the RPV operators appear simultaneously,
allowing observable RPV effects without disastrous proton decay.

Our goal in this paper is twofold.
Firstly, to present a detailed analysis of all possible combinations
of RPV operators arising from a generic semi-local F-theory spectral cover framework, assuming an $SU(5)$ GUT. 
This includes a detailed analysis of the classification of all possible allowed 
combinations of RPV operators, originating from 
the  $SU(5)$ term $10\cdot \bar 5\cdot \bar 5$,
including the effect of $U(1)$ fluxes, with {\em global} restrictions, which are crucial in
controlling the various possible multiplet splittings. Secondly, using F-theory techniques developed in the last few years, we 
perform explicit computations of the bottom/tau and RPV 
Yukawa couplings, assuming only {\em local} restrictions on fluxes, and 
comparing our results with the present experimental limits on the coupling for each specific RPV operator. 
The ingredients for this study have already appeared scattered through
the literature, which we shall refer to as we go along.

We emphasise that the first goal is related to the  nature of the available {\em global} Abelian fluxes of the particular model
 and their restrictions on the various matter curves, hence, on its specific geometric properties.  
The second goal  requires the computation of the strengths of the corresponding Yukawa couplings.
This in turn requires knowledge of the wavefunctions' profiles of the particles participating in
the corresponding trilinear Yukawa couplings and, as we will see, these involve the {\em local} flux 
data. Once such couplings exist in the  effective Lagrangian, we wish to explore the regions of the 
available parameter space where these  couplings are sufficiently suppressed and are compatible with the
present experimental data. 

Our aim in this dedicated study is to develop and extend the scope of the existing results in the literature, in order to 
provide a complete and comprehensive study, which
make direct contact with experimental limits on RPV, enabling F-theory models to be classified and
confronted with experiment more easily and directly than previously. 
We emphasise that this is the first study of its kind
in the literature which focusses exclusively on RPV in F-theory.

The remainder of the paper divides into two parts: in the first part, we consider semi-local
F-theory constructions where global restrictions are imposed on the fluxes, which imply that they 
take integer values. In Section~\ref{semi-local}
we show that RPV is a generic expectation of semi-local F-theory constructions.
In Section~\ref{curves} we classify 
F-theory $SU(5)$ models in the spectral cover approach according to the type of monodromy
which dictates the different curves on which the matter and Higgs fields can lie, with particular attention
of the possibility for RPV operators in each case at the level of $10\cdot \bar 5\cdot \bar 5$
operators, involving complete $SU(5)$ multiplets, focussing on which multiplets contain the Higgs fields
$H_u$ and $H_d$.
In Section~\ref{flux} we introduce the notion of flux, quantised according to global restrictions,
which, when switched on, leads to incomplete 
$SU(5)$ multiplets in the low energy (massless) spectrum, focussing on missing components of the multiplets
projected out by the flux, and tabulating the type of physical process (RPV or proton decay) can result
from particular operators involving different types of incomplete multiplets.
Appendix~\ref{app:RPV} details 
all possible sources of R-parity violating couplings for all
models classified with respect to the monodromies in semi-local F-theory constructions.

In the second part of the paper, we relax the global restrictions of the semi-local constructions, and allow the fluxes to take
general values, subject only to local restrictions.
In Section~\ref{yuk} we 
describe the calculation of a Yukawa coupling originating from 
an operator $10\cdot \bar 5\cdot \bar 5$ at an $SO(12)$ local point of enhancement in the presence of 
general local fluxes, with only local (not global) flux restrictions.
In Section~\ref{num} we apply these methods to calculate the numerical values of Yukawa couplings
for bottom, tau and RPV operators, exploring the parameter space of local fluxes.
In Section~\ref{RPV} we finally consider RPV coupling regions and
calculate ratios of Yukawa couplings from which the physical RPV couplings at the GUT scale 
can be determined and compared to limits on these couplings from experiment.
Section~\ref{conclusions} concludes the paper.
Appendix~\ref{app:LocalChirality} details the 
local F-theory constructions and local chirality constraints on flux data and RPV operators.

\section{R-parity violation in semi-local F-theory constructions}
\label{semi-local}
\subsection{Multi-curve models in the spectral cover approach}
\label{curves}

In the present F-theory framework of $SU(5)$ GUT,  third generation  fermion masses are expected to arise 
from  the  tree-level superpotential terms $10_f\cdot \bar 5_f\cdot \bar 5_{\bar H}$,  $10_f\cdot 10_f\cdot 5_H$
and $5_H\cdot \bar 5_{f}\cdot 1_f$, where the index $f$ stands for fermion, $H$ for Higgs
and we have introduced the notation
\be 
   10_f=(Q,u^c, e^c), \; \bar 5_f = (d^c,L),\; 1_f=\nu^c,\; 5_H=(D, H_u),\; \bar 5=(\bar D, H_d) \label{MSSMrpv}
\ee 
The  lighter generations receive masses from higher order terms,
involving the same invariants, although suppressed by powers of $\langle\theta_i\rangle/M$, with $\theta_i$
representing available singlet fields with non-zero vacuum expectation values (vevs), while $M$ is 
the GUT scale. 
 The 4-d RPV couplings are obtained similarly with the replacements $\bar 5_{\bar H}\to \bar 5_f$ 
(provided that the symmetries of the theory permit the existence of such terms).
At the level of the minimal supersymmetric standard model
(MSSM) superpotential the RPV couplings read~\cite{Barbier:2004ez}:
\begin{equation}\label{eq:WRPV}
W \supset 10_f\cdot \bar 5_f\cdot \bar 5_f \to  \mu_i H_u L_i+\frac{1}{2}\lambda_{ijk} L_i L_j e^c_k + \lambda^\prime_{ijk} L_i Q_j d^c_k + \frac{1}{2} \lambda^{\prime\prime}_{ijk}u^c_i d^c_j d^c_k
\end{equation}
in the conventional notation for matter multiplets $Q_i,u^c_i,d^c_i,L_i,e^c_i$ where $i=1,2,3$ is a flavour index. 
Notice that in the presence of vector-like pairs, $5_f+\bar 5_{f}$, additional RPV couplings 
appear from the following decompositions
\be 
\label{eq:WRPV2}
	W \supset 10_f\cdot 10_f\cdot 5_f \to \kappa Qu^c\bar L+\kappa'  u^c \bar d^c e^c+\frac 12\kappa'' QQ\bar d^c 
\ee 
where we have introduced the notation $5_f=(\bar d^c, \bar L)$ and dropped the flavour indices here for simplicity.
However, as we will analyse in detail, Abelian fluxes and additional continuous or discrete symmetries which are always
present in F-theory models, eliminate several of these terms. We will perform the analysis in the context of 
the spectral surfaces whose covering group is $SU(5)_{\perp}$ (dubbed usually as perpendicular) and 
is identified as the commutant to the GUT $SU(5) $ in the chain
\[ E_8\supset SU(5)\times SU(5)_{\perp}\to SU(5)\times U(1)^4_{\perp}   \]
where $E_8$ is assumed to be the highest singularity in the elliptically fibred compact space.
Then, a crucial r\^ole on the RPV remaining terms in the effective superpotential 
 is played by the specific assignment of fermion and Higgs fields on the various matter curves
and the remaining perpendicular  $U(1)_{\perp}$'s after  the monodromy action.   

A classification of the  set of models with simple monodromies that retain some perpendicular $U(1)_{\perp}$ charges
associated with the weights $t_i$
has been put forward in~\cite{Heckman:2009mn,Marsano:2009gv,Dudas:2010zb},
where we follow the notation of Dudas and Palti\cite{Dudas:2010zb} .
In the following, we categorize these models in order to assess whether tree-level, renormalizable, perturbative RPV 
is generic if matter is allocated in different curves. More specifically, we present four classes, characterised by the splitting of the spectral cover equation. These are:
\begin{itemize}
	\item
	$2+1+1+1$-splitting, which retains three independent perpendicular $U(1)_{\perp}$. These models represent a $Z_2$ monodromy ($t_1 \leftrightarrow t_2$), and as expected we are left with seven $\mathbf{5}$ curves, and four $\mathbf{10}$ curves.
	\item
	$2+2+1$-splitting, which retains two independent perpendicular $U(1)_{\perp}$. These models represent a  $Z_2\times Z_2$ monodromy ($t_1 \leftrightarrow t_2$, $t_3 \leftrightarrow t_4$),  and as expected we are left with five $\mathbf{5}$ curves, and three $\mathbf{10}$ curves.
	\item
	$3+1+1$-splitting, which retains two independent perpendicular $U(1)_{\perp}$. These models represent a  $Z_3$ monodromy ($t_1 \leftrightarrow t_2\leftrightarrow t_3$),  and as expected we are left with five $\mathbf{5}$ curves, and three $\mathbf{10}$ curves.
	\item
	$3+2$-splitting, which retains a single perpendicular $U(1)_{\perp}$. These models represent a $Z_3 \times Z_2$ monodromy ($t_1 \leftrightarrow t_2\leftrightarrow t_3$, $t_4 \leftrightarrow t_5$), and as expected we are left with three $\mathbf{5}$ curves, and two $\mathbf{10}$ curves.
\end{itemize}

In Appendix~\ref{app:RPV} we develop the above classes of models, identifying which curve contains the Higgs fields and which contains the matter fields, in order to show that RPV is a generic phenomenon
in semi-local F-theory constructions.
Of course, if all the RPV operators are present, then proton decay will be an inevitable consequence.
In the next subsection we show that this is generally avoided in semi-local F-theory constructions
when fluxes are switched on, which has the effect of removing some of the RPV operators,
while leaving some observable RPV in the low energy spectrum.

\subsection{Hypercharge flux with global restrictions and  R-parity violating operators}
\label{flux}

In F-theory GUTs, when the adjoint representation is not found in the massless spectrum,
the alternative  mechanism of flux breaking is introduced to reduce the GUT symmetry
down to the SM gauge group.  In the case of $SU(5)$ this can happen by turning on 
a non-trivial flux along the hypercharge generator in the internal directions. 
At the same time, the various components of the GUT multiplets living on 
matter curves, interact differently with the hypercharge flux. As a result,   
in addition to the  $SU(5)$ symmetry breaking, on certain matter curves we expect  
the splitting of the $10$ and $5,\bar 5$ representations  into different numbers of 
SM multiplets. 

In a minimal scenario one  might anticipate that the
hyperflux is non-trivially restricted only on the  Higgs matter curves in such a way that
the zero modes of the colour triplet components are eliminated. This would be 
an alternative to the doublet-triplet  scenario since only  the 
two Higgs doublets remain in the light spectrum. 
 The occurrence of this minimal 
scenario presupposes that all the other matter curves are left intact by the flux.
However, in this section we show that this is usually not the case.
Indeed,  the common characteristic of a large class of models derived from the various 
factorisations of the spectral cover are that there are incomplete $SU(5)$ multiplets
from different matter curves which comprise  the three known generations and eventually
possible extraneous fields. Interestingly, such scenarios leave open the possibility of 
effective models with only a fraction of RPV operators and the 
opportunity of studying  exciting new physics implications leading to 
suppressed exotic decays  which might be anticipated in the LHC experiments.

 To analyse these cases,  we assume  that  $m_{10}, m_{5}$ integers are units of $U(1)$ fluxes, with $n_Y$ 
  representing  the corresponding   hyperflux piercing  the matter curves.  
  The integer nature of these fluxes originates from the assumed {\em global} restrictions
  ~\cite{Heckman:2009mn,Marsano:2009gv,Dudas:2010zb}.
  Then,  the
tenplets and  fiveplets  split according to:
\be
    {10}_{t_{i}}=
                            \left\{\begin{array}{ll}{\rm
Representation}&{\rm flux\, units }\\
                         n_{{(3,2)}_{1/6}}-n_{{(\bar 3,2)}_{-1/6}}&=\;m_{10}\\
                        n_{{(\bar 3,1)}_{-2/3}}-n_{{(
3,1)}_{2/3}}&=\;m_{10}-n_Y\\
                        n_{(1,1)_{+1}}-n_{(1,1)_{-1}}& =\;m_{10}+n_Y\\
                            \end{array}\right.
                            \label{10s}
                            \ee
                            \be
                            {5}_{t_{i}}=
                            \left\{\begin{array}{ll}{\rm
Representation}&{\rm flux \, units }\\
                         n_{(3,1)_{-1/3}}-n_{(\bar{3},1)_{+1/3}}&=\;m_{5}\\
                        n_{(1,2)_{+1/2}}-n_{(1,2)_{-1/2}}& =\;m_{5}+n_Y\\
                            \end{array}\right.
                            \label{5s}
  \ee
The integers $m_{10,5}, n_Y$ may take any positive or negative value, leading to different numbers of SM representations, however, for our purposes it is enough to assume the cases \footnote{Of course there are several combinations of $(m,n_{Y})$ values which do not exceed the total number of three generations. Here, in order to illustrate the point,  we consider only the cases with $m,n_Y=\pm 1, 0$.}  $m,n_Y=\pm 1, 0$.
Then, substituting these numbers in Eqs.~(\ref{10s},\ref{5s}) we obtain the cases of Table~\ref{table1}.
\begin{table}[h!]
\centering\begin{tabular}{|c|c|c|c|c|c|}
\hline
{\bf $10$}&Flux units&\(10\) content&{\bf $\bar 5$}&Flux units &\(\bar
5\) content
\\
\hline
$10_1$&\(m_{10}=1, n_Y=0\) &$\{Q,u^c,e^c\}$ &$\bar 5_1$&\(m_{5}=1,
n_Y=0\)&$\{d^c,L\}$ \\
$10_2$&\(m_{10}=1, n_Y=1\) & \(\{Q,-,2 e^c\}\) &$\bar 5_2$&\(m_{5}=1,
n_Y=1\)&$\{d^c,2L\}$\\
$10_3$&\(m_{10}=1, n_Y=-1\) & \(\{Q,2u^c,-\}\) &$\bar 5_3$&\(m_{5}=1,
n_Y=-1\)&$\{d^c,-\}$\\
$10_4$&\(m_{10}=0, n_Y=1\) & \(\{-,\bar u^c,e^c\}\) &$\bar
5_4$&\(m_{5}=0, n_Y=1\)&$\{-,L\}$\\
$10_5$&\(m_{10}=0, n_Y=-1\) & \(\{-, u^c,\bar e^c\}\)  &$\bar
5_5$&\(m_{5}=0, n_Y=-1\)&$\{-,\bar L\}$\\
\hline
\end{tabular}
\caption{Table of MSSM matter content originating from $10,\ov{10}, 5,\bar 5$ of $SU(5)$ 
for various fluxes\label{table1}}
\end{table}
Depending on the specific choice of $m, n_Y$ integer parameters, we end up with incomplete $SU(5)$ representations.
For convenience we collect all distinct cases of incomplete $SU(5)$ multiplets in Table~\ref{table1}.

We now examine all parity violating operators formed by trilinear terms involving incomplete representations.
Table~\ref{table2} summarises the possible cases emerging form the various 
combinations $10_a\bar 5_b \bar 5_c$ of the incomplete representations shown in~Table  \ref{table1}.

\begin{table}[h!]\centering\begin{tabular}{|c|c|c|c|}
\hline
\( SU(5)\)-invariant&matter content&operators&Dominant
$\slash{\hspace{-.25cm} R}$-process
\\
\hline
$10_1\cdot \bar 5_1\cdot \bar 5_1$&$(Q,u^c,e^c)(d^c,L)^2$
&All&proton decay\\
$10_1\cdot \bar 5_2\cdot \bar 5_2$&\((Q,u^c,e^c)(d^c,2L)^2\)
&All&proton decay\\
$10_1\cdot\bar 5_3\cdot \bar 5_3$& \((Q,u^c,e^c)(d^c,-)^2\)&$u^cd^cd^c$&$n-\bar
n$-osc.\\
$10_1\cdot \bar 5_4\cdot \bar 5_4$& \((Q,u^c,e^c)(-,L)^2\)
&$LL e^c$&$L_{e,\mu,\tau}$-violation\\
$10_1\cdot \bar 5_5\cdot \bar 5_5$& \((Q,u^c,e^c)(-,\bar L)^2\)
&None&None\\
\hline
$10_2\cdot \bar 5_1\cdot \bar 5_1$&$(Q,-,e^c)(d^c,L)^2$ &$QL
d^c,LL e^c$&$L_{e,\mu,\tau}$-violation\\
$10_2\cdot \bar 5_2\cdot \bar 5_2$&\((Q,-,e^c)(d^c,2L)^2\) &$QL
d^c,LL e^c$&$L_{e,\mu,\tau}$-violation\\
$10_2\cdot\bar 5_3\cdot \bar 5_3$& \((Q,-,e^c)(d^c,-)^2\)&None&None\\
$10_2\cdot \bar 5_4\cdot \bar 5_4$& \((Q,-,e^c)(-,L)^2\) &$LL
e^c$&$L_{e,\mu,\tau}$-violation\\
$10_2\cdot \bar 5_5\cdot \bar 5_5$& \((Q,-,e^c)(-,\bar L)^2\)  &None&None\\
\hline
$10_3\cdot \bar 5_1\cdot \bar 5_1$&\((Q,2 u^c,-)(d^c,L)^2\) &$QL
d^c,d^cd^c u^c$&proton decay\\
$10_3\cdot \bar 5_2\cdot \bar 5_2$&\((Q,2 u^c,-)(d^c,2L)^2\)
&$QL d^c,d^cd^c u^c$&proton decay\\
$10_3\cdot\bar 5_3\cdot \bar 5_3$& \((Q,2 u^c,-)(d^c,-)^2\)&$d^cd^c
u^c$&$n-\bar n$-osc.\\
$10_3\cdot \bar 5_4\cdot \bar 5_4$& \((Q,2 u^c,-)(-,L)^2\) &None&None\\
$10_3\cdot \bar 5_5\cdot \bar 5_5$& \((Q,2 u^c,-)(-,\bar L)^2\)
&None&None\\
\hline
\end{tabular}
\caption{Fluxes, incomplete representations and
$\slash{\hspace{-.25cm} R}$-processes emerging from the trilinear coupling 
$10_a\bar 5_b \bar 5_c$ for all possible 
combinations of the incomplete  multiplets given in Table~\ref{table1}. \label{table2}}
\end{table}

In the last column of Table~\ref{table2} we also show the dominant RPV processes, which lead to baryon and/or lepton number violation. We notice however, that there exist other rare processes
beyond those indicated in the tables   which can be found in reviews (see for example~\cite{Barbier:2004ez}.)
We have already stressed, that  in addition to the standard model particles, some vector-like pairs may appear too. For example,
 when fluxes are turned on, we have seen in several cases that the MSSM spectrum is accompanied in vector like states such as:
 
  \[ u^c+\bar u^c, L+\bar{L}, d+\ov{d}^c, Q+\ov{Q}\ldots  \]          
  Of course they are expected to get a heavy mass but if some vector-like pairs remain in the light spectrum  
  they  may have significant implications in rare processes, such as contributions to diphoton events which are
  one of the primary searches in the ongoing LHC experiments.

\section{Yukawa couplings in local F-theory constructions: formalism}
\label{yuk}

In this section (and subsequent sections) we relax the global constraints on fluxes,
and consider the calculation of Yukawa couplings, imposing only local
flux restrictions. The motivation for doing this is to calculate the Yukawa couplings 
associated with the RPV operators in a rather model independent way, 
and then compare our results to the experimental limits.
Flavour hierarchies and Yukawa structures in F-theory have been studied in a large number 
of papers \cite{Heckman:2008qa}-\cite{Hebecker:2014uaa}. In this section we shall discuss 
Yukawa couplings in F-theory, following the approach of \cite{Font:2009gq, Aparicio:2011jx,Font:2012wq}.

In the previous section we assessed how chirality is realised on different curves due to flux effects. These considerations take into account the global flux data and are therefore called semi-local models. The flux units considered in the examples above are integer valued as they follow from the Dirac flux quantisation
\begin{equation}\label{eq:GlobalFlux}
	\frac{1}{2 \pi} \int_{\Sigma \subset S} F = n
\end{equation}
where $n$ is an integer, $\Sigma$ a matter curve (two-cycle in the divisor $S$), and $F$ the gauge  field-strength tensor, i.e. the flux. In conjugation with the index theorems, the flux units piercing different matter curves $\Sigma$ will tell us how many chiral states are globally present in a model.

While the semi-local approach defines the full spectrum of a model, the computation of localised quantities, such as the Yukawa couplings, requires appropriate description of the local geometry. As we will see below, a crucial quantity in the local geometry is the notion of {\em local} flux density, understood as follows.

First we notice that the unification gauge coupling is related to the compactification scale through the volume of the compact space
\begin{equation}
	\alpha_G^{-1} = m^4_* \int_S 2 \omega \wedge \omega = m^4_* \int {\rm d Vol}_S = \text{Vol}(S) m^4_*
\end{equation}
where $\alpha_G$ is the unification gauge coupling, $m_*$ is the F-Theory characteristic mass, $S$ the GUT divisor with K\"ahler form
\begin{equation}
\omega = \frac{i}{2}( dz_1 \wedge  d\bar z_1+ dz_2 \wedge  d\bar z_2)
\end{equation}
that defines the volume form
\begin{equation}
	{\rm d Vol}_S = 2 \omega \wedge \omega = dz_1 \wedge dz_2 \wedge d\bar z_1 \wedge d\bar z_2.
\end{equation}
As the volume of $\Sigma$ is bounded by the volume of $S$, we assume that
\begin{equation}
	\text{Vol}(\Sigma) \simeq \sqrt{\text{vol}(S)} , 
\end{equation}
and if we now consider that the background of $F$ is constant, we can estimate the values that $F$ takes in $S$ by
\begin{equation}
	F \simeq 2 \pi \sqrt{\alpha_G} m^2_* n .
\end{equation}
This means that, in units of $m_*$, the background $F$ is an $\mathcal{O}(1)$ real number. Since in the computation of Yukawa couplings it's the local values of $F$ -- and not the global quantisation constraints -- that matter, we will from now on abuse terminology and refer to flux densities, $F$, as fluxes. Furthermore, as we will see later, the local values of $F$ also define what chiral states are supported locally. This will be crucial to study the full plenitude of RPV couplings in different parts of the parameter space.

Before dealing with the particular rare reaction, it is useful to recall a few basic facts about the Yukawa couplings.

\subsection{The local $SO(12)$ model}

In F-theory matter is localised along Riemann surfaces (matter curves), which are formed at the intersections of D7-branes with the GUT surface $S$. Yukawa couplings are then realised when three of these curves intersect at a single point on $S$, while, at the same time, the gauge symmetry is enhanced. The computation relies on the knowledge of the profile of the wavefunctions of the states participating in the intersection. When a specific geometry is chosen for the internal space (and in particular for the GUT surface) these profiles are found by solving the corresponding equations of motion~\cite{Cecotti:2009zf}-\cite{Font:2012wq}. Their values are obtained  by  computing the integral of the overlapping wavefunctions  at the triple intersections.

In $SU(5)$ two basic Yukawa terms are relevant when computing the Yukawa matrices and interactions. These are  $\mathbf{y_u 10\cdot 10\cdot 5}$  and  $\mathbf{y_d 10\cdot \bar 5\cdot \bar 5}$. The first one generates the top Yukawa coupling while the symmetry at this intersection enhances to the exceptional group $E_6$.  The relevant couplings that we are interested in, are related to the second coupling. This one is realised at a point where there is an $SO(12)$ gauge symmetry enhancement\footnote{For  a general $E_8$ point of enhancement that containing both type of couplings see \cite{Palti:2012aa,Marchesano:2015dfa}. Similar, an $E_7$ analysis is given in \cite{Carta:2015eoh}.}. To make this clear, next we highlighted some of the basic analysis of \cite{Font:2012wq}.

The 4-dimensional theory can be obtained by integrating out the effective 8-dimensional one over the divisor $S$
\begin{equation}
W = m_*^4 \int_S \text{Tr} (F\wedge \Phi)
\end{equation}
where $F=dA-iA\wedge{A}$ is the field-strength of the gauge vector boson $A$ and $\Phi$ is a $(2,0)$-form on $S$.

From the above superpotential, the F-term equations can be computed by varying $A$ and $\Phi$. In conjugation with the D-term
\begin{equation}
	D = \int_S \omega \wedge F + \frac{1}{2} [\Phi,\bar \Phi], 
\end{equation}
where $\omega$ is the K\"ahler form of $S$, a 4-dimensional supersymmetric solution for the equations of motion of $F$ and $\Phi$ can be computed.

Both $A$ and $\Phi$, locally are valued in the Lie algebra of the symmetry group at the Yukawa point. In the case in hand, the fibre develops an $SO(12)$ singularity at which point couplings of the form ${\bf 10}\cdot\bar{\bf 5}\cdot\bar{\bf 5}$ arise. Away from the enhancement point, the background $\Phi$ breaks $SO(12)$ down to the GUT group $SU(5)$. The r\^{o}le of $\langle A \rangle$ is to provide a 4d chiral spectrum and to break further the GUT gauge group.

More systematically, the Lie-Algebra of $SO(12)$ is composed of its Cartan generators $H_i$ with $i=1,...,6$, and $60$ step generators $E_\rho$. Together, they respect the Lie algebra
\begin{equation}
[H_i, E_\rho] = \rho_i E_\rho
\end{equation}
where $\rho_i$ is the $i^{th}$ component of the root $\rho$. The $E_\rho$ generators can be completely identified by their roots
\begin{equation}
(\underline{\pm 1 , \pm 1 , 0 , 0, 0, 0, 0})
\end{equation}
where underline means all 60 permutations of the entries of the vector, including different sign combinations.
To understand the meaning of this notation it is sufficient to consider a simpler example:
\begin{equation}
(0, \underline{1 , 0 , 0}, 0, 0, 0) \equiv \{ (0,1,0,0,0,0  ), (0,0,1,0,0,0  ),(0,0,0,1,0,0  )\}
\end{equation}

The background of $\Phi$ will break $SO(12)$ away from the $SO(12)$ singular point. In order to see this consider it takes the form
\begin{equation}
\Phi = \Phi_{z_1z_2}  dz_1 \wedge dz_2
\end{equation}
where it's now explicit that it parametrises the transverse directions to $S$. The background we are considering is
\begin{equation}\label{eq:BackgroundHiggs}
\langle \Phi_{z_1z_2} \rangle = m^2 \left(z_1 Q_{z_1}+ z_2 Q_{z_2}\right)
\end{equation}
where $m$ is related to the slope of the intersection of 7-branes, and
\begin{align}
Q_{z_1} &= - H_{1} \\
Q_{z_2} &= \frac{1}{2} \sum_i H_i .
\end{align}

The unbroken symmetry group will be the commutant of $\langle \Phi_{z_{1}z_{2}} \rangle$ in $SO(12)$. The commutator between the background and the rest of the generators is
\begin{equation}
[\langle\Phi_{z_1z_2}\rangle,E_\rho] = m^2 q_{\Phi}(\rho) E_\rho
\end{equation}
where $q_\Phi(\rho)$ are holomorphic functions of the complex coordinates $z_1$, $z_2$. The surviving symmetry group is composed of the generators that commute with $\langle \Phi \rangle$ on every point of $S$. With our choice of background, the surviving step generators are identified to be
\begin{equation}
E_\rho : (0,\underline{1,-1,0,0,0}),
\end{equation}
which, together with $H_i$,  trivially commute with $\langle \Phi \rangle$, generating $SU(5) \times U(1) \times U(1)$. 

When $q_\Phi(\rho)=0$ in certain loci we have symmetry enhancement, which accounts for the presence of matter curves. This happens as at these loci, extra step generators survive and furnish a representation of $SU(5)\times U(1) \times U(1)$. For the case presented we identify three curves joining at the $SO(12)$ point, these are
\begin{align}
\Sigma_a &= \{z_1=0\} \\
\Sigma_b &= \{z_2=0\} \\
\Sigma_c &= \{z_1=z_2\} ,
\end{align}
and defining a charge under a certain generator as
\begin{equation}\label{key}
[Q_i, E_\rho] = q_i(\rho) E_\rho
\end{equation}
all the data describing these matter curves are presented in Table \ref{tab:SO12curves}. Since the bottom and tau Yukawas come from such an $SO(12)$ point, in order to have such a coupling the point must have the $a^+$, $b^+$, and $c^+$.

\begin{table}
\begin{center}
{\renewcommand{\arraystretch}{1.5}
	\begin{tabular}{|c|c|c|c|c|c|}
	\hline 
	Curve  & Roots  & $q_\Phi$  & $SU(5)$ irrep  & $q_{z_1}$  & $q_{z_2}$  \\ 
	\hline 
	$\Sigma_{a^{\pm}}$& $(\pm 1, \underline{\mp 1, 0, 0, 0, 0})$  & $\mp z_1$  & $\bar{\bf 5}/{\bf 5}$  & $\mp 1$  & 0  \\ 
	$\Sigma_{b^{\pm}}$& $(0,\underline{\pm 1, \pm 1, 0, 0, 0})$ & $\mp z_2$  & ${\bf 10}/\bar{\bf 10}$  & $0$  & $\pm 1$  \\ 
	$\Sigma_{c^{\pm}}$& $(\mp 1, \underline{\mp 1, 0, 0, 0, 0})$ & $\pm (z_1-z_2)$  & $\bar{\bf 5}/{\bf 5}$  & $\pm 1 $  & $\mp 1$  \\ 
	\hline 
\end{tabular}}
\end{center}
\caption{\label{tab:SO12curves} Matter curves and respective data for an $SO(12)$ point of enhancement model with a background Higgs given by Equation \ref{eq:BackgroundHiggs}. The underline represent all allowed permutations of the entries with the signs fixed}
\end{table} 	

In order to both induce chirality on the matter curves and break the two $U(1)$ factors, we have to turn on fluxes on $S$ valued along the two Cartan generators that generate the extra factors.

We first consider the flux 
\begin{equation}
\langle F_1 \rangle = i (M_{z_1} d{z_1} \wedge d\bar {z_1} + M_{z_2} d{z_2}\wedge d\bar {z_2}) Q_F ,
\end{equation}
with
\begin{equation}
Q_F = -Q_{z_1} - Q_{z_2} = \frac{1}{2} (H_1 - \sum_{j=2}^{6}H_{j}).
\end{equation}

It's easy to see that the $SU(5)$ roots are neutral under $Q_F$, and therefore this flux does not break the GUT group. On the other hand, the roots on $a$, $b$ sectors are not neutral. This implies that this flux will be able to differentiate  $\bar{\bf 5}$ from ${\bf 5}$ and ${\bf 10}$ from $\bar {\bf 10}$
\begin{equation}
\int_{\Sigma_a, \ \Sigma_b} F_1 \neq 0 \Rightarrow \mbox{ Induced Chirality} . 
\end{equation}

This flux does not induce chirality in $c^\pm$ curves as $q_F=0$ for all roots in $c^\pm$. To induce chirality in $c^\pm$ one needs another contribution to the flux
\begin{equation}
\langle F_2 \rangle = i (d{z_1}\wedge d\bar {z_2}+d{z_2} \wedge d\bar {z_1})(N_a Q_{z_1} + N_b Q_{z_2})
\end{equation}
that does not commute with the roots on the $c^\pm$ sectors for $N_a \neq N_b$.

Breaking the GUT down to the SM gauge group requires flux along the Hypercharge. In order to avoid generating a Green-Schwarz mass for the Hypercharge gauge boson, this flux has to respect global constraints. Locally we may define it as
\begin{equation}
\langle F_Y \rangle =i [(d{z_1}\wedge d\bar {z_2} + d{z_2} \wedge d \bar {z_1})N_Y + (d{z_2} \wedge d\bar {z_2} - d {z_1} \wedge d \bar {z_1})\tilde N_Y] Q_Y
\end{equation}
and the Hypercharge is embedded in our model through the linear combination
\begin{equation}
Q_Y = \frac{1}{3} (H_2+H_3+H_4) - \frac{1}{2}(H_5+H_6) .
\end{equation}

Since this contribution to the flux does not commute with all elements of $SU(5)$, only with its SM subgroup, distinct SM states will feel this flux differently. This known fact is used extensively in semi-local models as a mechanism to solve the doublet-triplet splitting problem. As we will see bellow, it can also be used to locally prevent the appearance of certain chiral states and therefore forbid some RPV in subregions of the parameter space.

The total flux will then be the sum of the three above contributions. It can be expressed as
\begin{align}
\langle F \rangle =& i (d{z_2} \wedge d \bar {z_2} - d {z_1} \wedge d \bar {z_1}) Q_P\nonumber \\
&+ i (d{z_1}\wedge d\bar {z_2} + d {z_2} \wedge d \bar {z_1} ) Q_S \nonumber\\
&+ i (d{z_2}\wedge d\bar{z_2} + d{z_1} \wedge d\bar {z_1})M_{{z_1}{z_2}} Q_F
\end{align}
with the definitions
\begin{align}
Q_P = &  M Q_F + \tilde N_Y Q_Y \\
Q_S = &   N_a Q_{z_1} + N_b Q_{z_2} + N_Y Q_Y
\end{align}
\noindent and
\begin{align}
M   =  &  \frac{1}{2}(M_{z_1}-M_{z_2})\\
M_{z_{1}z_{2}} =& \frac{1}{2} (M_{z_2} + M_{z_1}) . 
\end{align}

As the Hypercharge flux will affect SM states differently, breaking the GUT group, we will be able to distinguish them inside each curve. The full split of the states present in the different sectors, and all relevant data, is presented in Table \ref{tab:SO12curvesWflux}.

\begin{table}
	\begin{center}
	{\renewcommand{\arraystretch}{1.5}
		\begin{tabular}{ | c | c | c  | c | c | c | c| c|}		\hline
		Sector & Root                                    & SM           & $q_F$ & $q_{z_1}$ & $q_{z_2}$ & $q_S$                    & $q_P$                           \\ \hline
		$a_1$  & $(1,\underline{-1,0,0},0,0)$            & $(\bar{\bf 3}, {\bf 1})_{-\frac{1}{3}}$ & $1$   & $-1$  & $0$   & $-N_a -\frac{1}{3} N_Y$  & $M -\frac{1}{3} \tilde{N}_Y$    \\
		$a_2$  & $(1,0,0,0,\underline{-1,0})$            & $({\bf 1}, {\bf 2})_{\frac{1}{2}}$         & $1$   & $-1$  & $0$   & $-N_a+\frac{1}{2}N_Y$    & $M +\frac{1}{2} \tilde{N}_Y$    \\ 
		$b_1$  & $(0,\underline{1,1,0},0,0)$             & $(\overline{\bf 3},{\bf 1})_{\frac{2}{3}}$   & $-1$  & $0$   & $1$   & $N_b+\frac{2}{3}N_Y$     & $- M +\frac{2}{3} \tilde{N}_Y$  \\ 
		$b_2$  & $(0,\underline{1,0,0},\underline{1,0})$ & $({\bf 3},{\bf 2})_{-\frac{1}{6}}$             & $-1$  & $0$   & $1$   & $N_b-\frac{1}{6}N_Y$     & $- M - \frac{1}{6} \tilde{N}_Y$ \\ 
		$b_3$  & $(0,0,0,0,1,1)$                         & $({\bf 1},{\bf 1})_{-1}$                       & $-1$  & $0$   & $1$   & $N_b-N_Y$                & $ - M - \tilde{N}_Y$            
		\\ 
		$c_1$  & $(-1,\underline{-1,0,0},0,0)$           & $(\bar{\bf  3},{\bf 1})_{-\frac{1}{3}}$      & $0$   & $1$   & $-1$  & $N_a-N_b-\frac{1}{3}N_Y$ & $-\frac{1}{3} \tilde{N}_Y$      \\ 
		$c_2$  & $(-1,0,0,0,\underline{-1,0})$           & $({\bf 1},{\bf 2})_{\frac{1}{2}}$              & $0$   & $1$   & $-1$  & $N_a-N_b+\frac{1}{2}N_Y$ & $\frac{1}{2} \tilde{N}_Y$       \\ \hline
	\end{tabular}}
\end{center}
	\caption{\label{tab:SO12curvesWflux}Complete data of sectors present in the three curves crossing in an $SO(12)$ enhancement point considering the effects of non-vanishing fluxes. The underline represent all allowed permutations of the entries with the signs fixed}
\end{table}

\subsection{Wavefunctions and the Yukawa computation}

In general, the Yukawa strength is obtained by computing the integral of the overlapping wavefunctions. More  precisely, according to the discussion on the previous section one has to solve for the zero mode wavefunctions for the sectors $a,b$ and $c$ presented in Table (\ref{tab:SO12curvesWflux}). The physics of the D7-Branes wrapping on $S$ can be described in terms of a twisted 8-dimensional $\mathcal{N}=1$ gauge theory on $R^{1,3}\times{S}$, where $S$ is a K\"{a}hler submanifold of elliptically fibered Calabi-Yau 4-fold $X$. One starts with the action of the effective theory, which was given in  \cite{Beasley:2008kw}. The next step is to obtain the equations of motion for the 7-brane fermionic zero modes. This procedure has been performed in several of papers including~\cite{Palti:2012aa,Aparicio:2011jx,Font:2012wq} and we will not repeat it here in detail. 
In order for this paper to be  self-contained we highlight the basic computational steps.

The equations for a 4-dimensional massless fermionic field are of the Dirac form:

\begin{equation}\label{eom1}
\mathcal{D}_{A}\Psi=0
\end{equation}

\noindent where

\begin{equation}\label{eom2}
\mathcal{D}_{A} =\left(
\begin{array}{cccc}
0 & D_{1} & D_{2} & D_3 \\
-D_{1} & 0 & -D_{\bar{3}} & D_{\bar{2}} \\
-D_{2} & -D_{\bar{3}}& 0& -D_{\bar{1}} \\
-D_{3} & -D_{\bar{2}} & D_{\bar{1}} & 0
\end{array}\right)\quad{,}\quad{\Psi=\Psi E_{\rho}=\left(\begin{array}{c}
-\sqrt{2}\eta\\ \psi_{\bar{1}}\\ \psi_{\bar{2}}\\ \chi_{12}
\end{array}\right).}
\end{equation}

The indices  here are a shorthand notation instead of the coordinates $z_{1},z_{2},z_{3}$. The components of $\Psi$ are representing 7-brane degrees of freedom. Also the covariant derivatives are defined as $D_{i}=\partial_{i}-i[\langle A_{i}\rangle,\ldots]$ for $i=1,2,\bar{1},\bar{2}$ and as $D_{\bar{3}}=-i[\langle \Phi_{12}\rangle,\ldots]$ for the coordinate $z_3$. It is clear from equations (\ref{eom1},\ref{eom2}) that we have to solve the equations for each sector. According to the detailed solutions in \cite{Font:2012wq} the wavefunctions for each sector have the general form

\begin{equation}
\Psi\sim{f(az_{1}+bz_{2})e^{M_{ij}z_{i}z_{j}}}
\end{equation}

\noindent where $f(az_{1}+bz_{2})$ is a holomorphic function and $M_{ij}$ incorporates flux effects. In an appropriate basis this holomorphic function can be written as a power of its variables  $f_{i}\sim{(az_{1}+bz_{2})}^{3-i}$ and in the case where the generations reside in the same matter curve, the index-$i$ can play the r\^{o}le of a family index. Moreover the Yukawa couplings as a triple wavefunction integrals have to respect geometric $U(1)$ selection rules. The coupling must be invariant under geometric transformations of the form: $z_{1,2}\rightarrow{e^{i\alpha}z_{1,2}}$. In this case the only non-zero tree level coupling arises for $i=3$ and by considering that, the  index in the holomorphic function $f_i$ indicates the fermion generation  we obtain a non-zero top-Yukawa coupling. Hierarchical couplings for the other copies  on the same matter curve can be generated in the presence of non commutative fluxes \cite{Cecotti:2009zf} or by incorporating  non-perturbative effects \cite{Aparicio:2011jx}-\cite{Carta:2015eoh}.

\begin{figure}[t]
	\centering
	\includegraphics[scale=0.5]{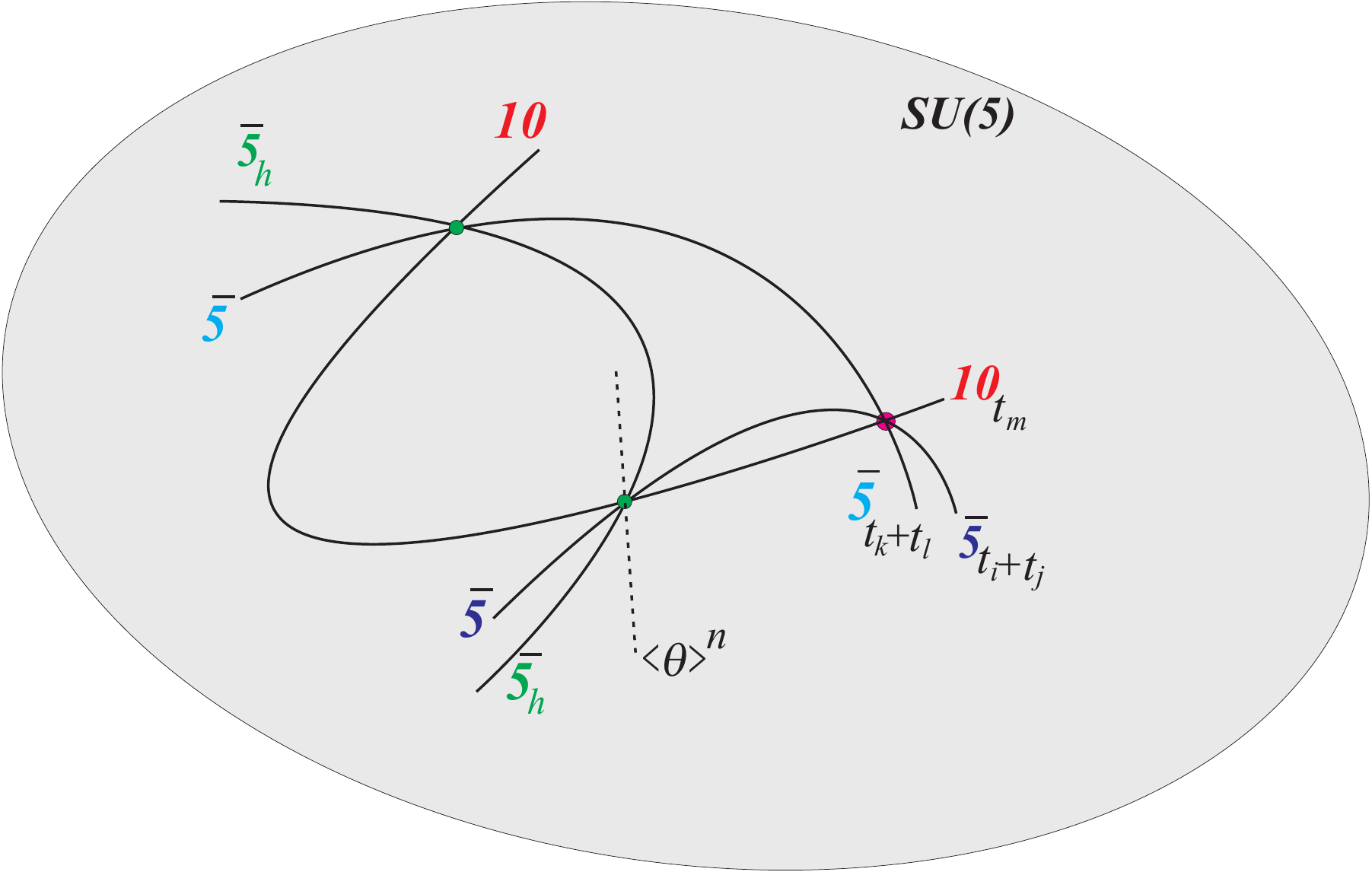}
	\caption{Intersecting matter curves, Yukawa couplings and the case of RPV .}\label{cartoon}
\end{figure}

The RPV couplings under consideration emerge from a  tree level interaction. Hence, its strength is given by computing  the
integral  where now the r\^{o}le of the Higgs $\bar 5_{H}$ is replaced by $\bar 5_M$.   We consider here the scenario where the generations are 
accommodated in different matter curves. In this case the two couplings, the bottom/tau Yukawa and the tree level RPV, are  localised at different $SO(12)$ points on $\mathcal{S}_{GUT}$, (see Figure \ref{cartoon}). In this approach, at first approximation we can take the holomorphic functions $f$ as constants absorbed in the normalization factors.

 As a first approach, our goal is to calculate the bottom Yukawa coupling as well as the coupling without hypercharge flux and compare the two values.
So, at this point we  write down the wavefunctions and the relevant parameters  in a more detailed form as given in \cite{Font:2012wq} but without the holomorphic functions. The wavefunctions in the holomorphic gauge  have the following form
\begin{align}
	\vec{\psi}_{_{10_{M}}}^{(b)hol} & =\vec{v}^{(b)}\chi_{_{10_{M}}}^{(b)hol}=\vec{v}^{(b)}\kappa^{(b)}_{_{10_{M}}}e^{\lambda_{b}z_{2}(\bar{z}_{2}-\zeta_{b}\bar{z}_{1})}\label{eq:psi10M}                   \\
	\vec{\psi}_{_{5_{M}}}^{(a)hol}  & =\vec{v}^{(a)}\chi_{_{5_{M}}}^{(a)hol}=\vec{v}^{(a)}\kappa^{(a)}_{_{5_{M}}} e^{\lambda_{a}z_{1}(\bar{z}_{1}-\zeta_{a}\bar{z}_{2})}\label{eq:psi5M}                       \\
	\vec{\psi}_{_{5_{H}}}^{(c)hol}  & =\vec{v}^{(c)}\chi_{_{5_{H}}}^{(c)hol}=\vec{v}^{(c)}\kappa_{_{5_{H}}}^{(c)} e^{(z_{1}-z_{2})(\zeta_{c}\bar{z}_{1}-(\lambda_{c}-\zeta_{c})\bar{z}_{2})}\label{eq:psi5H}   \\
	\vec{\psi}_{_{5_{M}}}^{(c)hol}  & =\vec{v}^{(c)}\chi_{_{5_{H}}}^{(c)hol}=\vec{v}^{(c)}\kappa_{_{5_{M}}}^{(c)} e^{(z_{1}-z_{2})(\zeta_{c}\bar{z}_{1}-(\lambda_{c}-\zeta_{c})\bar{z}_{2})}\label{eq:psi5Mc}.
\end{align}
 where
\begin{align}
	\zeta_{a} &= - \frac{ q_S(a)}{\lambda_a -q_P(a)}\\
	\zeta_{b} &= - \frac{ q_S(b)}{\lambda_b+q_P(b)}\\
	\zeta_{c} &= \frac{\lambda_c (\lambda_c -q_P(c)-q_S(c)}{2(\lambda_c-q_S(c))}
\end{align}
and $\lambda_\rho$ is the smallest eigenvalue of the matrix
\begin{equation}
m_\rho =\left(
\begin{array}{ccc}
-{q_P} & {q_S} & i {m}^2 {q_{z_1}} \\
{q_S} & {q_P} & i {m}^2 {q_{z_2}} \\
-i {m}^2 {q_{z_1}} & -i {m}^2 {q_{z_2}} &
0 \\
\end{array}
\right).
\end{equation}

To compute the above quantities we make use of the values of $q_i$ from Table \ref{tab:SO12curvesWflux}. It is important to note that the values of the flux densities in this table  depend on the $SO(12)$ enhancement point. This means that one can in principle have different numerical values for the strength of the interactions at different points.

The column vectors are given by

 \begin{equation}\label{columns}
\quad \vec{v}^{(b)}=\left(\begin{array}{c}
	-\frac{i\lambda_{b}}{m^2}\zeta_{b} \\
	         \frac{i\lambda_{b}}{m^2}          \\
	                        1
\end{array}\right),\vec{v}^{(a)}=\left(\begin{array}{c} -\frac{i\lambda_{a}}{m^2}\\\frac{i\lambda_{a}}{m^2}\zeta_{a}\\1\\\end{array}\right), \vec{v}^{(c)}=\left(\begin{array}{c} -\frac{i\zeta_{c}}{m^2}\\\frac{i(\zeta_{c}-\lambda_{c})}{m^2}\\1\\\end{array}\right).
\end{equation}

 \noindent Finally, the $\kappa$ coefficients in equations (\ref{eq:psi10M}-\ref{eq:psi5M})  are normalization factors. These factors are fixed by imposing canonical kinetic terms for the matter fields. More precisely, for a canonically normalized field $\chi_i$ supported in a certain sector $(e)$,  the normalization condition for the wavefunctions in the real gauge is
\begin{equation}
1 =2 m_*^4 ||\vec v^{(e)} ||^2 \int (\chi^{(e)real})_i^* \chi^{(e)real}_i {\rm d Vol}_S
\end{equation} 
where $\chi_i^{(e)real}$ are now in the real gauge, and in our convention $\mbox{Tr}E^\dagger_\alpha E_\beta = 2 \delta_{\alpha \beta}$. The wavefunctions in real and holomorphic gauge are related by
\begin{equation}
	\psi^{real} = e^{i\Omega} \psi^{hol}
\end{equation}
where
\begin{equation}
	\Omega = \frac{i}{2} \left[ \left(M_{z_1} |z_1|^2 + M_{z_2} |z_2|^2\right)Q_F - \tilde N_Y \left(|z_1|^2-|z_2|^2\right)Q_Y + \left(z_1 \bar z_2 + z_2 \bar z_1\right)Q_S \right],
\end{equation}
which only transforms the scalar coefficient of the wavefunctions, $\chi$, leaving the $\vec{v}$ part invariant.

With the above considerations, one can find the normalization factors to be
\begin{equation}
|\kappa_{5_{M}}^{(a)}|^{2}=-4\pi g_{s}\sigma^{2}\cdot\frac{q_{P}(a)(2\lambda_{a}+q_{P}(a)(1+\zeta_{a}^{2}))}{\lambda_{a}(1+\zeta_{a}^{2})+m^{4}},\label{norm5M}
\end{equation} 
  \begin{equation}
|\kappa_{10_{M}}^{(b)}|^{2}=-4\pi g_{s} \sigma^{2}\cdot\frac{q_{P}(b)(-2\lambda_{b}+q_{P}(b)(1+\zeta_{b}^{2}))}{\lambda_{b}(1+\zeta_{b}^{2})+m^{4}},\label{norm10M} 
\end{equation}
   \begin{equation}
|\kappa_{5_{H}}^{(c)}|^2=-4\pi g_{s} \sigma^{2}\cdot\frac{2(q_{P}(c)+\zeta_{c})(q_{P}(c)+2\zeta_{c}-2\lambda_{c})+(q_{S}(c)+\lambda_{c})^{2}}{\zeta_{c}^{2}+(\lambda_{c}-\zeta_{c})^{2}+m^{4}},\label{norm5H} 
\end{equation}
   \begin{equation}
   |\kappa_{5_{M}}^{(c)}|^{2}=-4\pi g_{s} \sigma^{2}\cdot\frac{2(q_{P}(c)+\zeta_{c})(q_{P}(c)+2\zeta_{c}-2\lambda_{c})+(q_{S}(c)+\lambda_{c})^{2}}{\zeta_{c}^{2}+(\lambda_{c}-\zeta_{c})^{2}+m^{4}}.\label{norm5Mc} 
   \end{equation}

\noindent where we used the relation $
 \left(\frac{m}{m_*}\right)^2 = (2\pi)^{3/2} g_s^{1/2}\sigma$, making use of the dimensionless quantity $\sigma=(m/m_{st})^{2}$, where $m_{st}$ the string scale. The expressions (\ref{norm5M}-\ref{norm5Mc}) above can be shown numerically to be always positive.

 The superpotential trilinear couplings can be taken to be in the holomorphic gauge. For the bottom Yukawa, we consider that $\psi_{_{10_{M}}}$ and $\psi_{5_{M}}$ contain the heaviest down-type quark generations. In this case the bottom and tau couplings can be computed:
 
\begin{align}
y_{_{b, \tau}}=&m_{*}^{4}\  t_{abc}\int_{S}\det(\vec{\psi}^{(b)hol}_{_{10_{M}}},\vec{\psi}^{(a)hol}_{_{5_{M}}},\vec{\psi}^{(c)hol}_{_{5_{H}}}){\rm d Vol}_{S}\nonumber \\
=&m_{*}^4\  t_{abc}\  \det(\vec{v}^{(b)},\vec{v}^{(a)},\vec{v}^{(c)})\int_{S}\chi^{(b)hol}_{_{10_{M}}}\chi^{(a)hol}_{_{5_{M}}}\chi^{(c)hol}_{_{5_{H}}}{\rm d Vol}_{S}.\label{ybottom}
\end{align}
The bottom and tau Yukawa couplings differ since they have different SM quantum numbers and arise from different
sectors, leading to different $q_S$ and $q_P$ as shown in Table~\ref{tab:SO12curvesWflux}.

\noindent  A similar formula can be written down for the RPV coupling

\begin{align}
y_{_{RPV}}=&m_{*}^{4}\ t_{abc}\int_{S}\det(\vec{\psi}^{(b)hol}_{_{10_{M}}},\vec{\psi}^{(a)hol}_{_{5_{M}}},\vec{\psi}^{(c)hol}_{_{5_{M}'}}){\rm d Vol}_{S}\nonumber \\
=&m_{*}^4\ t_{abc}\  \det(\vec{v}^{(b)},\vec{v}^{(a)},\vec{v}^{(c)})\int_{S}\chi^{(b)hol}_{_{10_{M}}}\chi^{(a)hol}_{_{5_{M}}}\chi^{(c)hol}_{_{5_{M}}}{\rm d Vol}_{S}.\label{yRPV}
\end{align}
Here this RPV Yukawa coupling can in principle refer to any generations of squarks and sleptons,
and may have arbitrary generation indices (suppressed here for simplicity).
 
 \noindent The factor $t_{abc}$ represents the structure constants of the $SO(12)$ group.  The integral in the last term can be computed by applying standard Gaussian techniques. Computing the determinant and the integral, the combined result of the two is a flux independent factor and the final result reads:

 \begin{equation}
y_{_{b, \tau}}=\pi^{2}\left(\frac{m_{*}}{m}\right)^{4}t_{abc}\kappa_{_{10_{M}}}^{(b)}\kappa_{_{5_{M}}}^{(a)}\kappa_{_{5_{H}}}^{(c)}.\label{bottomfinal}
\end{equation}

 \noindent This is a standard result for the heaviest generations. As we observe the flux dependence is hidden on the normalization factors.

 We turn now our attention in the case of a tree-level RPV coupling of the form $10_{M}\cdot \bar{5}_{M}\cdot \bar{5}_{M}$. This coupling can be computed in a different $SO(12)$ enhancement point $p$. As a first approach we consider that the hypercharge flux parameters are zero in the vicinity of $p$. From a different point of view, $\bar{5}_{M}$ replaces the Higgs matter curve in the previous computation. The new wavefunction ($\psi^{(c)}_{_{5_{M}}}$) can be found by setting all the Hypercharge flux parameters on $\psi^{(c)}_{_{5_{H}}}$, equal to zero. The RPV coupling will be  given by an equation similar to that of the bottom coupling :
 
  \begin{equation}
 y_{_{RPV}}=\pi^{2}\left(\frac{m_{*}}{m}\right)^{4}t_{abc}\kappa_{_{10_{M}}}^{(b)}\kappa_{_{5_{M}}}^{(a)}\kappa_{_{5_{M}}}^{(c)}.\label{rpvfinal}
 \end{equation}
 and we notice that family indices are understood and this coupling is the same for every type of RPV interaction, depending on which SM states are being supported at the $SO(12)$ enhancement point. Notice that the $\kappa$'s in equations (\ref{bottomfinal}, \ref{rpvfinal}) are 
the modulus of the normalization factors defined in equations (\ref{norm5M}-\ref{norm5Mc}).

In the next section, using equations \eqref{bottomfinal} and \eqref{rpvfinal}, we perform a numerical analysis for the couplings presented above with emphasis on the case of the RPV coupling. We notice that in our conventions for the normalization of the $SO(12)$ generators, the gauge invariant coupling supporting the above interactions has $t_{abc}=2$·.

\section{Yukawa couplings in local F-theory constructions: numerics}
\label{num}
Using the mathematical machinery developed in the previous section, we can study the behaviour of $SO(12)$ points in F-theory - including both the bottom-tau point of enhancement and RPV operators. The former has been well studied in \cite{Font:2012wq} for example. The coupling is primarily determined by five parameters - $N_a$, $N_b$, $M$, $N_Y$ and $\tilde{N_Y}$. The parameters $N_a$ and $N_b$ give net chirality to the $c$-sector, while $N_Y$ and $\tilde{N_Y}$ are components of hypercharge flux, parameterising the doublet triplet splitting. $M$ is related to the chirality of the $a$ and $b$-sectors. There is also the $N_b = N_a - \frac{1}{3} N_Y $ constraint, which ensures the elimination of Higgs colour triplets at the Yukawa point. This can be seen by examining the text of the previous section, based on the work found in \cite{Font:2012wq}.\\

For a convenient and comprehensive presentation of the results we make the following redefinitions. In Eq. (\ref{bottomfinal}) and (\ref{rpvfinal}), one can factor out $4\pi g_{s}\sigma^{2}$ from inside Eq. (\ref{norm5M}),(\ref{norm10M}), and (\ref{norm5H}). In addition by noticing that  
$\left(\frac{m}{m_*}\right)^2 = (2\pi)^{3/2} g_s^{1/2}\sigma$,
we obtain
\begin{align}
	y_{_{b,\tau}} &=2 g_s^{1/2} \sigma \ y_{_{b, \tau }}^\prime \label{eq:ybprime}\\
	y_{_{RPV}} &= 2g_s^{1/2} \sigma \ y_{_{RPV}}^\prime \label{eq:yRPVprime} 
\end{align}
where $y^\prime_{b, \tau}$ and $y^\prime_{RPV}$ are functions of the flux parameters. Furthermore, we set the scale $m=1$ and as such the remainder mass dimensions are given in units of $m$. The presented values for the strength of the couplings are then in units of $2 g_s^{1/2}\sigma$.

\begin{figure}[h!]
\centering
\includegraphics[width=1\linewidth]{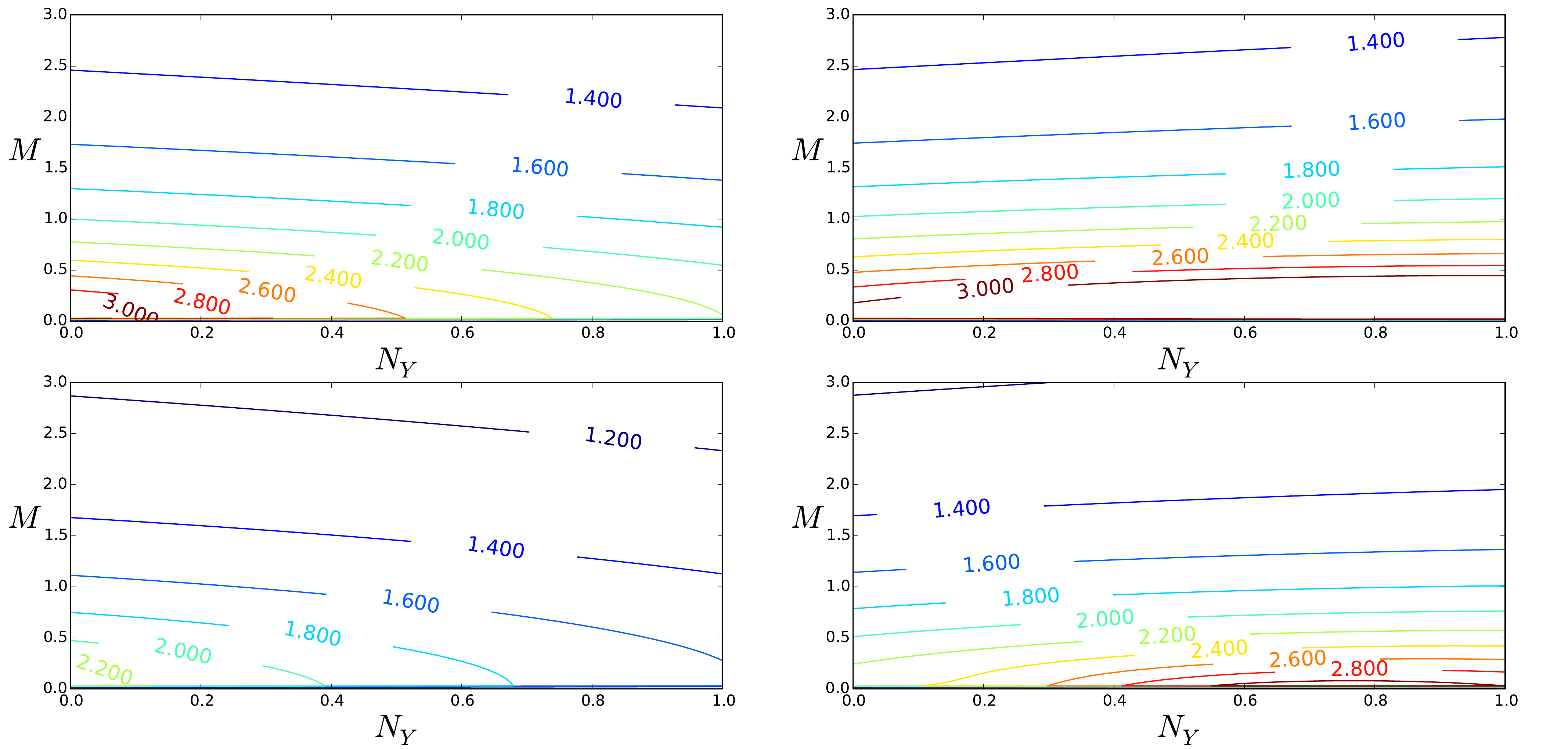}
\caption{Ratio between bottom Yukawa and tau Yukawa couplings, shown as contours
in the plane of local fluxes. The requirement for chiral matter and absence of coloured Higgs triplets fixes $N_b = N_a - \frac{1}{3} N_Y $}
\label{fig:BTratio}
\end{figure}

\fref{fig:BTratio} shows the ratio of the bottom and tau Yukawa couplings at a point of $SO(12)$ in a region of the parameter space with reasonable values. 
These results are consistent with those in~\cite{Font:2012wq}.
Note that the phenomenological desired 
ratio of the couplings at the GUT scale is $Y_{\tau}/Y_b=1.37\pm0.1\pm0.2$ \cite{Ross:2007az},
which can be achieved within the parameter ranges shown in \fref{fig:BTratio}.
Having shown that this technique reproduces the known results for the bottom to tau ratio,
we now go on to study the behaviour of an RPV coupling point in $SO(12)$ models.

\subsection{Behaviour of $SO(12)$ points}

\begin{figure}[t!]
\centering
\includegraphics[width=1\linewidth]{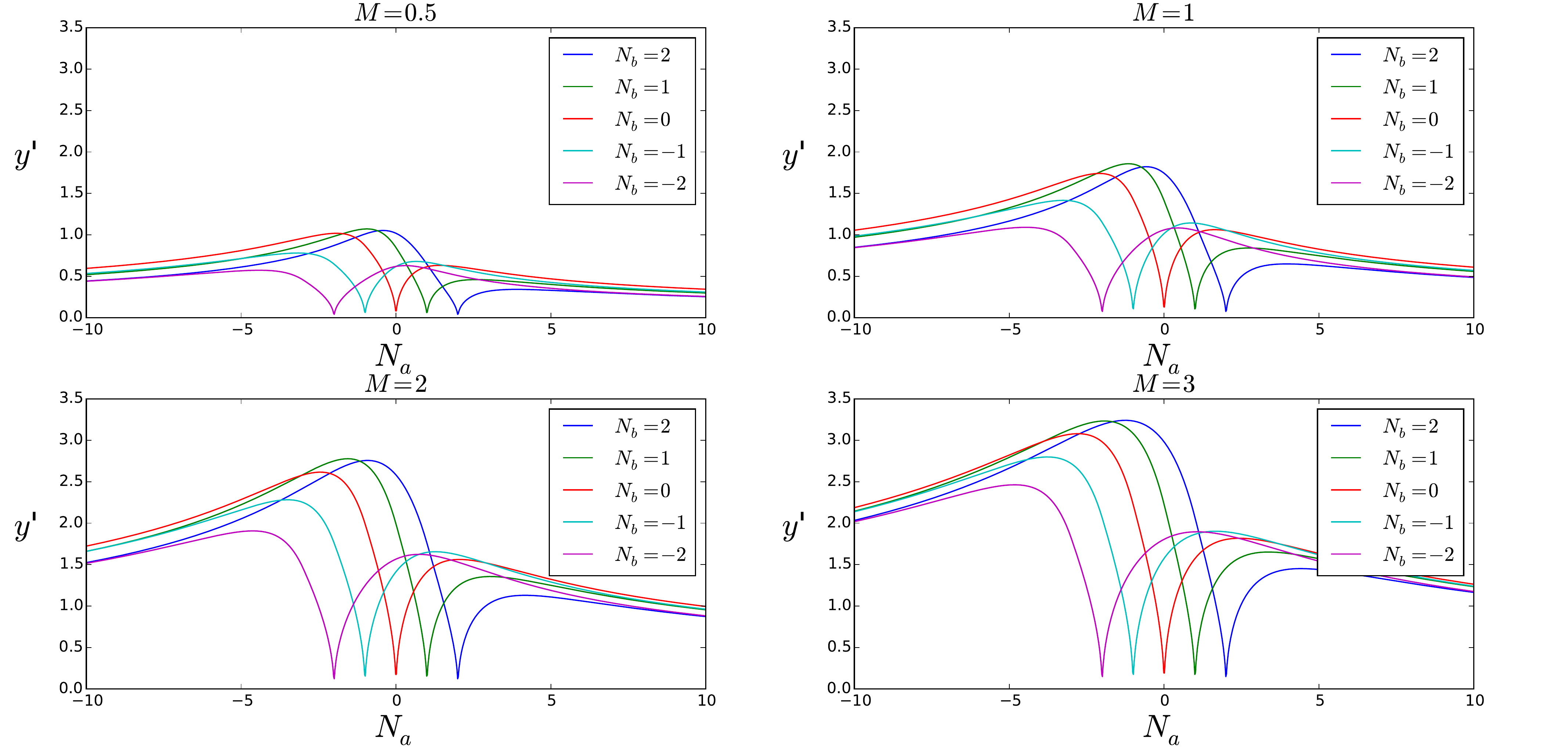}
\caption{Dependency of the RPV coupling (in units of $2 g_s^{1/2}\sigma$) on $N_a$ in the absence of hypercharge fluxes, for different values of $M$ and $N_b$.}
\label{fig:RpvNbvalsMvals}
\end{figure}

The simplest scenario for an $SO(12)$ enhancement generating RPV couplings, would be the case where all three of the types of operator, $QLD$, $UDD$, and $LLE$ arise with equal strengths, which would occur in a scenario with vanishing hypercharge flux, leading to an entirely \textquotedblleft unsplit" scenario. This assumption sets $N_Y$ and $\tilde{N_Y}$ to vanish, and we may also ignore the condition $N_b = N_a - \frac{1}{3} N_Y $.  The remaining parameters determining are then $N_a$, $N_b$ and $M$. \fref{fig:RpvNbvalsMvals} shows the coupling strength in the $N_a$ plane for differing $N_b$ and $M$ values. The general behaviour is that the coupling strength is directly related to $M$, while the coupling vanishes at the point where $N_a=N_b$. This latter point is due to the flip in net chirality for the $c$-sector at this point in the parameter space - $N_a>N_b$ gives the $c^+$ part of the spectrum.

\begin{figure}[h!]
	\centering
	\includegraphics[width=1\linewidth]{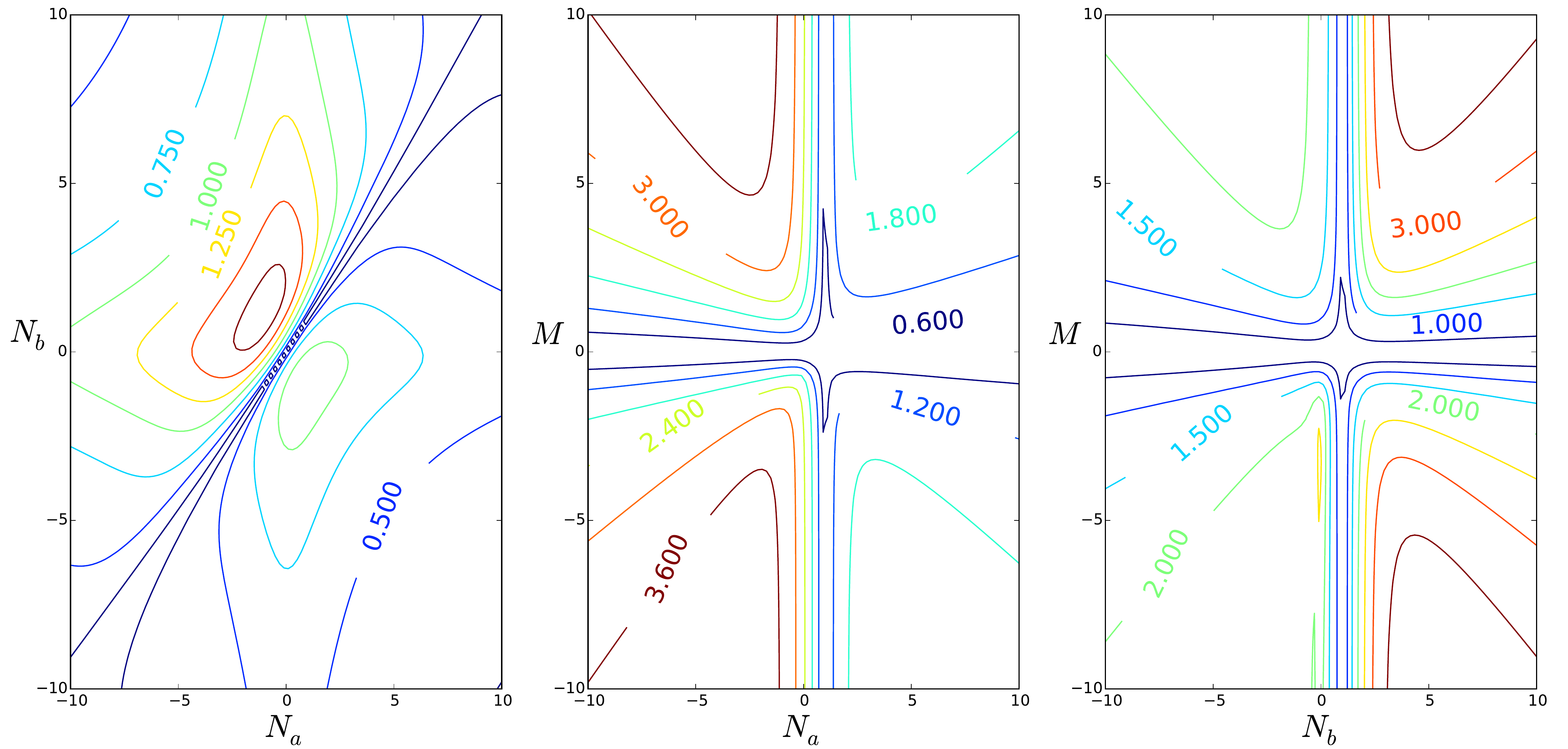}
	\caption{Dependency of the RPV coupling (in units of $2 g_s^{1/2}\sigma$) on different flux parameters, in absence of Hypercharge fluxes. Any parameter whose dependency is not shown is set to zero.}
	\label{fig:RPVFlux}
\end{figure}

\begin{figure}[h!]
	\centering
	\includegraphics[width=1\linewidth]{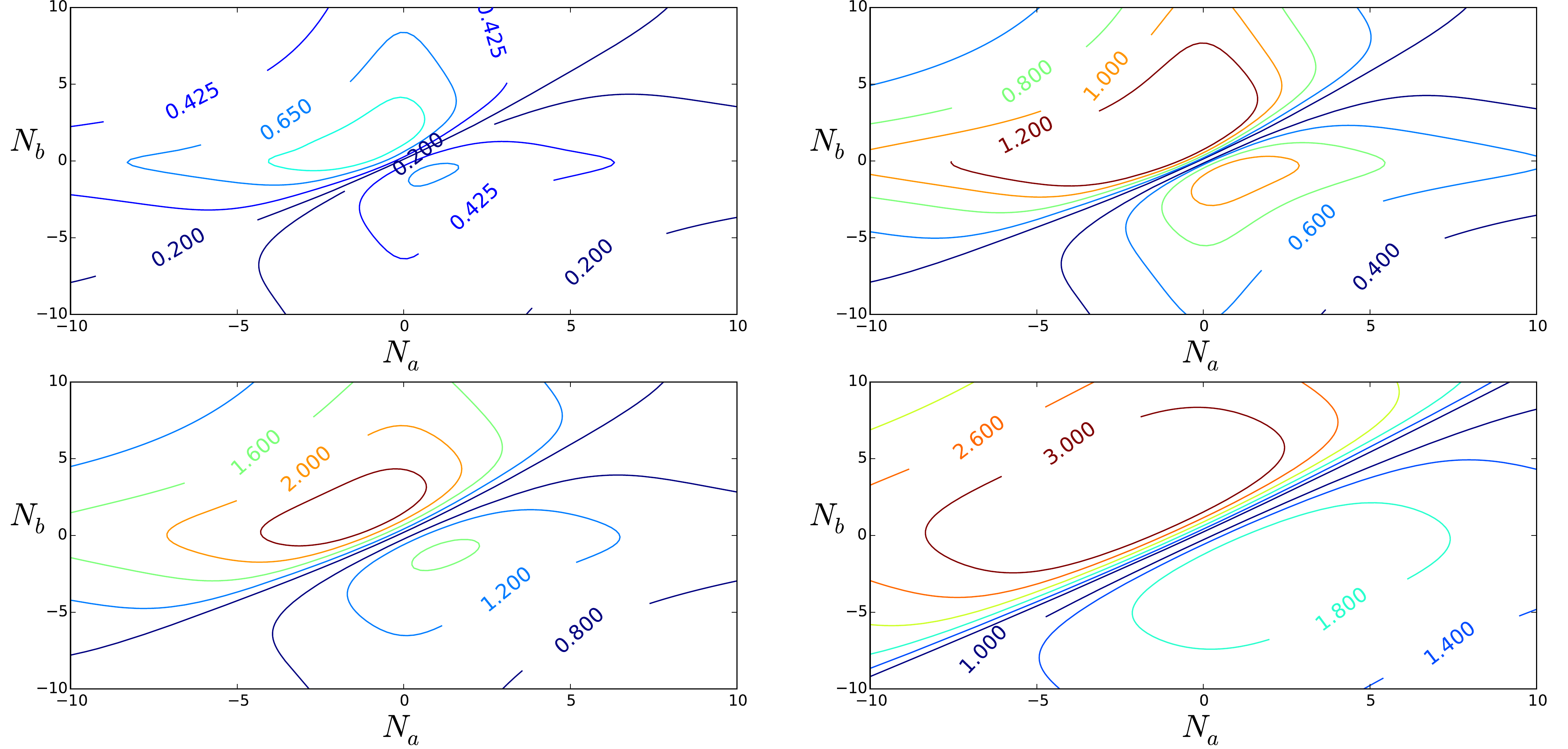}
	\caption{Dependency of the RPV coupling (in units of $2 g_s^{1/2}\sigma$) on the $(N_a,N_b)$-plane, in absence of hypercharge fluxes and for different values of $M$. Top: left $M=0.5$, right $M=1.0$. Bottom: left $M=2.0$, right $M=3.0$.}
	\label{fig:RPVNaNbM}
\end{figure}

\fref{fig:RPVFlux} and \fref{fig:RPVNaNbM} also demonstrate this set of behaviours, but for contours of the coupling strength. \fref{fig:RPVFlux}, showing all combinations of the three non-zero parameters, shows that in the $N_a-N_b$ plane there is a line of vanishing coupling strength about the $N_a=N_b$, chirality switch point for the $c$-sector. The figure also reinforces the idea that small values of $M$ correspond to small values of the coupling strength, as close to the point of $M=0$ the coupling again reduces to zero. \fref{fig:RPVNaNbM} again shows this behaviour, with the smallest values of $M$ giving the smallest values of the coupling. From this we can infer that an RPV $SO(12)$ point is most likely to be compatible with experimental constraints if $M$ takes a small value.

\begin{figure}[h!]
	\centering
	\subfigure[Varying $N_a$ with fixed $M$\label{fig:RpvB-Na}]
	{\includegraphics[width=0.325\linewidth]{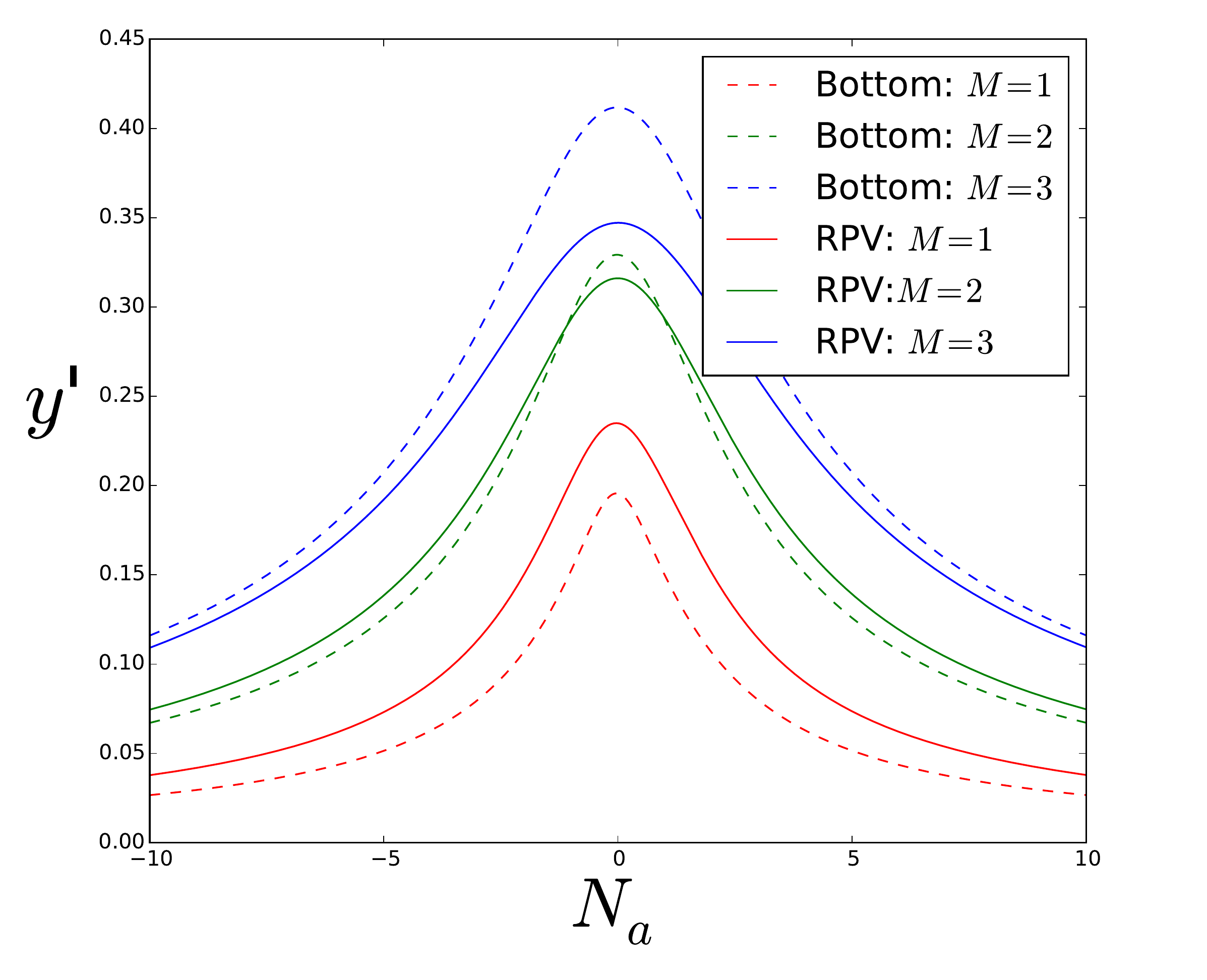}}
	\subfigure[Varying $N_b$ with fixed $M$\label{fig:RpvB-Nb}]
	{\includegraphics[width=0.325\linewidth]{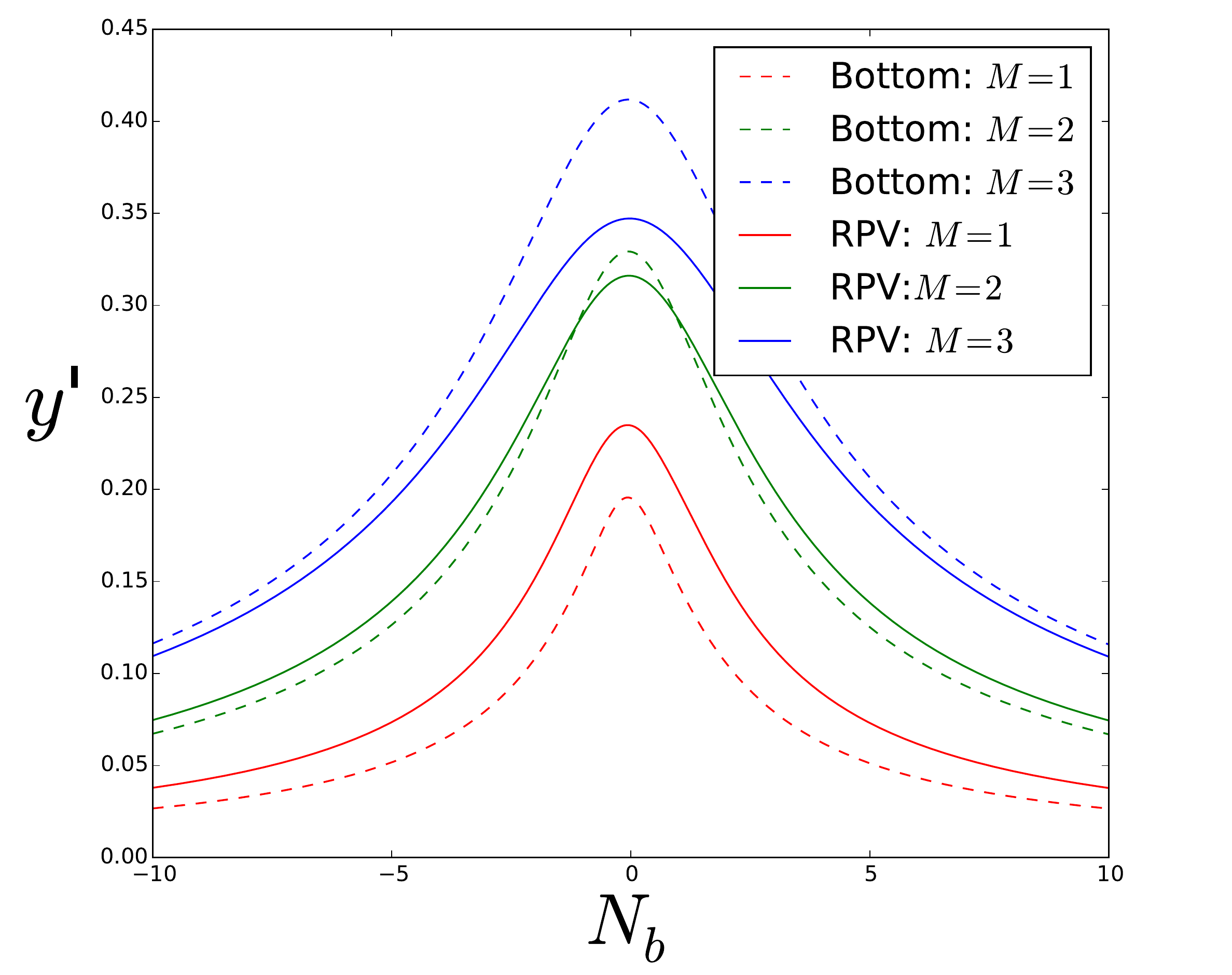}}
	\subfigure[Varying $M$ with fixed $N_a$=1, $N_b= 29/30$, $N_Y=0.1$, $\tilde N_Y = 3.6$\label{fig:RpvB-M}]
	{\includegraphics[width=0.325\linewidth]{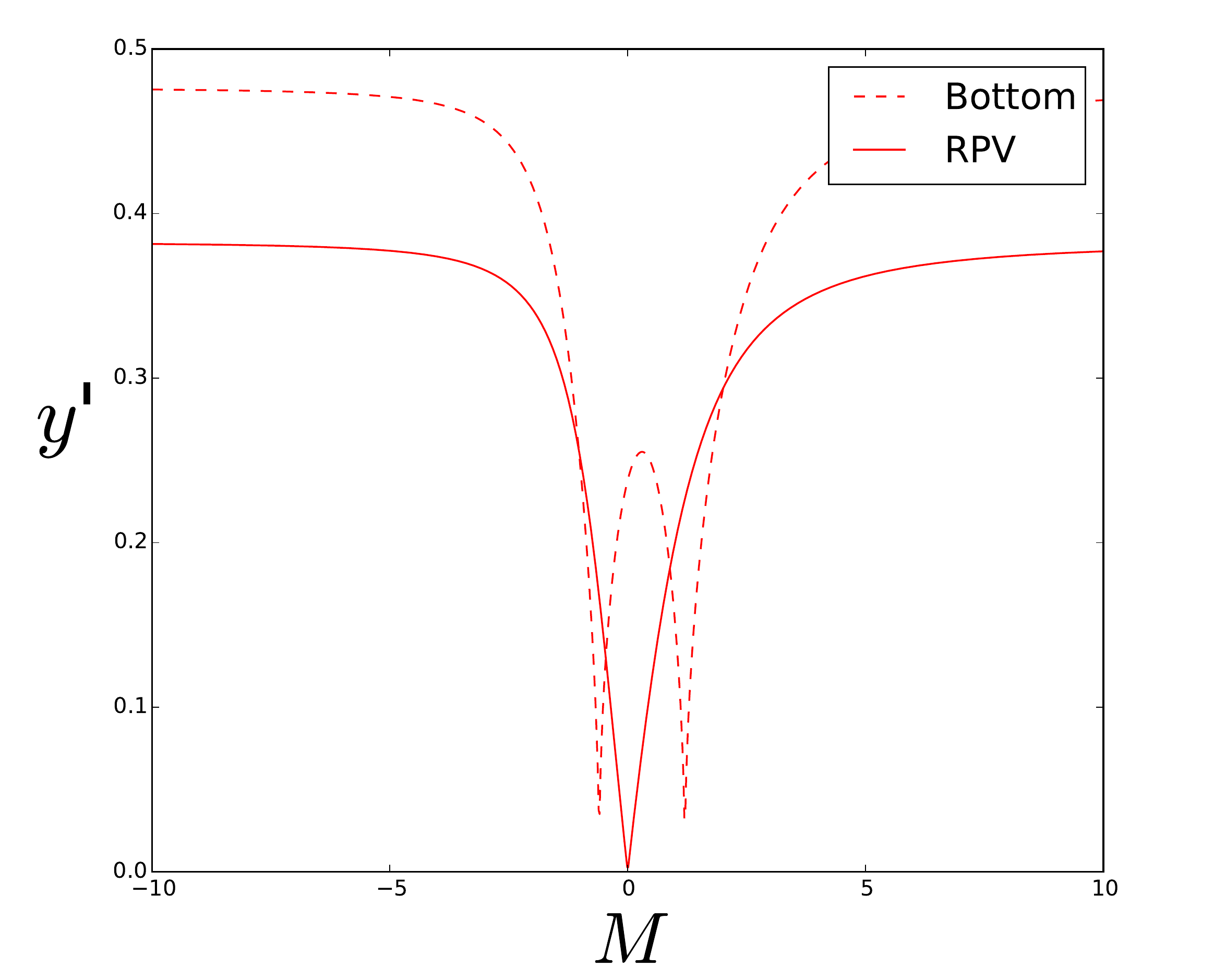}}
	\caption{Dependency of the RPV and bottom Yukawa couplings (in units of $2 g_s^{1/2}\sigma$) on different parameters at different regions of the parameter space}
\end{figure}

\fref{fig:RpvB-Na} (and \fref{fig:RpvB-Nb}) shows the RPV coupling strength in the absence of flux for the $N_a$ ($N_b$) plane, along with the \textquotedblleft bottom" coupling strength for corresponding values. The key difference is that the Hypercharge flux  is switched on at the bottom $SO(12)$ point, with values of  $N_Y=0.1$ and $\tilde N_Y = 3.6$. The figures show that for the bottom coupling, the fluxes always push the coupling higher, similarly to increasing the $M$ values. 

\fref{fig:RpvB-M} plots out the two couplings in the $M$-plane, showing that the bottom Yukawa goes to zero for two values of $M$, while the RPV point has only one. Considering the form of \eref{bottomfinal}, we can see that the factors $\kappa_{5_M}$ and $\kappa_{10_M}$ are proportional to the parameter $q_p$. Referring to \tref{tab:SO12curvesWflux}, one can see which values these take for each sector - namely, $q_p(a_1)=M-\frac{1}{3}\tilde{N_Y}$ and $q_p(b_2)=-M-\frac{1}{6}\tilde{N_Y}$. Solving these two equations shows trivially that zeros should occur when $M=\frac{1}{3}\tilde{N_Y}$ and $-\frac{1}{6}\tilde{N_Y}$, which is the exact behaviour exhibited in \fref{fig:RpvB-M}.

\section{R-parity violating Yukawa couplings: allowed regions and comparison to data}
\label{RPV}

In this section we focus on calculating the RPV Yukawa coupling constant at the GUT scale,
which may be directly compared to the experimental limits,
using the methods and results of the previous two sections.
As a point of notation, we have denoted the RPV Yukawa coupling at the GUT scale to be generically
$y_{RPV}$, independently of flavour or operator type indices. This coupling may be directly compared
to the phenomenological RPV Yukawa couplings at the GUT scale $\lambda_{ijk}$, $ \lambda^\prime_{ijk}$ and $\lambda^{\prime\prime}_{ijk}$
as defined below.

Recall that, in the weak/flavour basis, the superpotential generically includes RPV couplings, in particular
those from Eq. \ref{eq:WRPV}:
\begin{equation}
	W \supset \frac{1}{2}\lambda_{ijk} L_i L_j e^c_k + \lambda^\prime_{ijk} L_i Q_j d^c_k + \frac{1}{2} \lambda^{\prime\prime}_{ijk}u^c_i d^c_j d^c_k
\end{equation}
In the local F-theory framework, each of the above Yukawa couplings (generically denoted as
$y_{RPV}$) is computable through Eq. (\ref{rpvfinal}). What distinguishes different RPV couplings, say 
$\lambda$ from $\lambda^\prime$, are the values of the flux densities, namely the hypercharge flux. This is because the normalization of matter curves depends on the hypercharge flux density. As such, different SM states will have different hypercharges and consequently different respective normalization coefficient.

\begin{figure}[h!]
\centering
\includegraphics[width=0.7\linewidth]{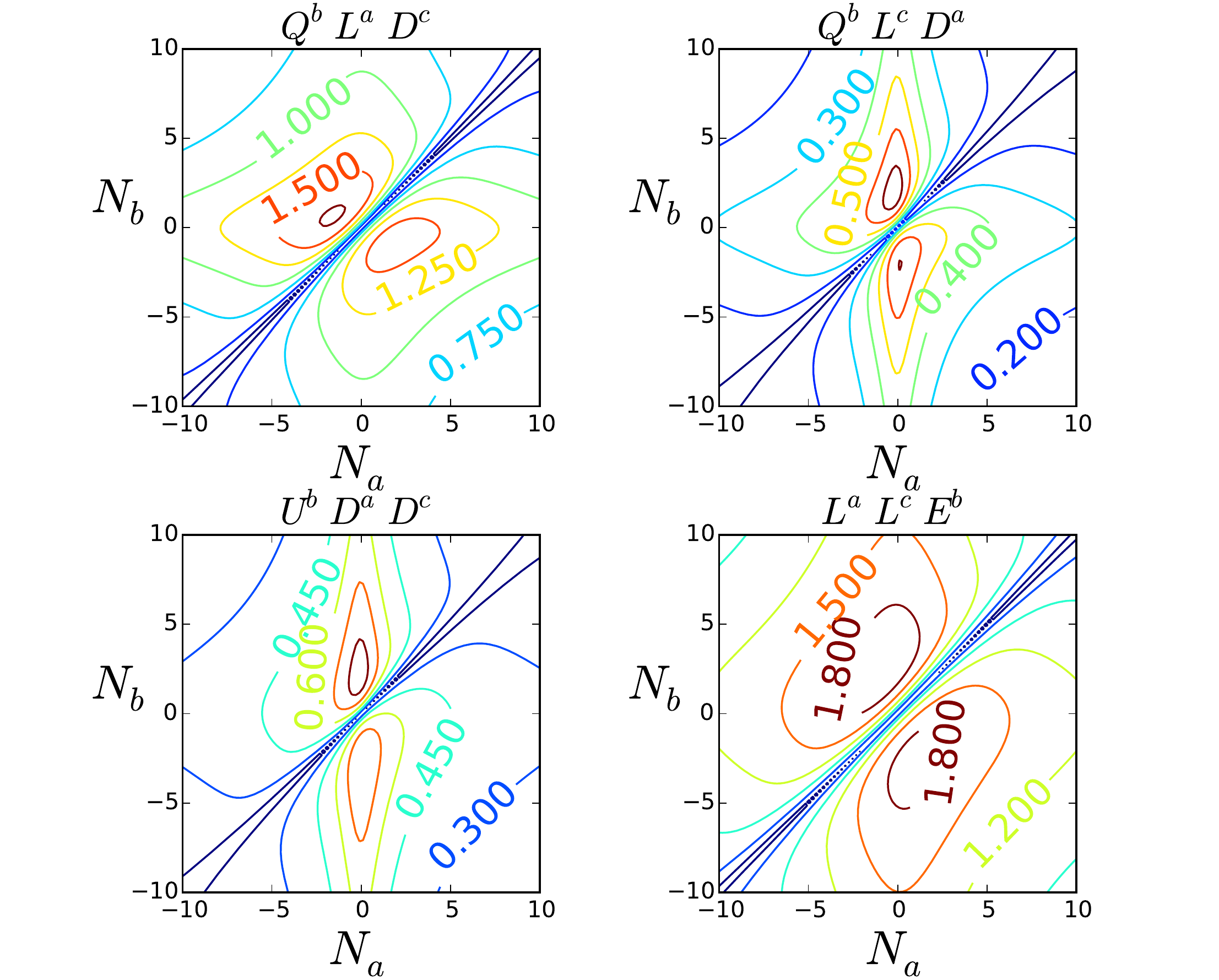}
\caption{Strength of different RPV couplings (in units of $2 g_s^{1/2}\sigma$) in the $(N_a,N_b)$-plane in the presence of Hypercharge fluxes $N_Y=0.1$, $\tilde N_Y=3.6$, and with $M=1$. The scripts $a$, $b$, $c$ refer to which sector each state lives.}
\label{fig:split-rpv-NaNb}
\end{figure}

Even though a given $SO(12)$ enhancement point can in principle support different types of trilinear RPV interactions, the actual effective interactions arising at such point depend on the local chiral spectrum present at each curve. For example, in order to have an $LLe^c$ interaction, both $\Sigma_a$ and $\Sigma_c$ curves need to have chiral $L$ states, and the $\Sigma_b$ curve an $e^c$ state at the enhancement point. In Figure \ref{fig:split-rpv-NaNb} we show contours on the ($N_a$,$N_b$) plane for the different types of trilinear RPV couplings. 

\begin{figure}[h!]
\centering
\includegraphics[width=1.0\linewidth]{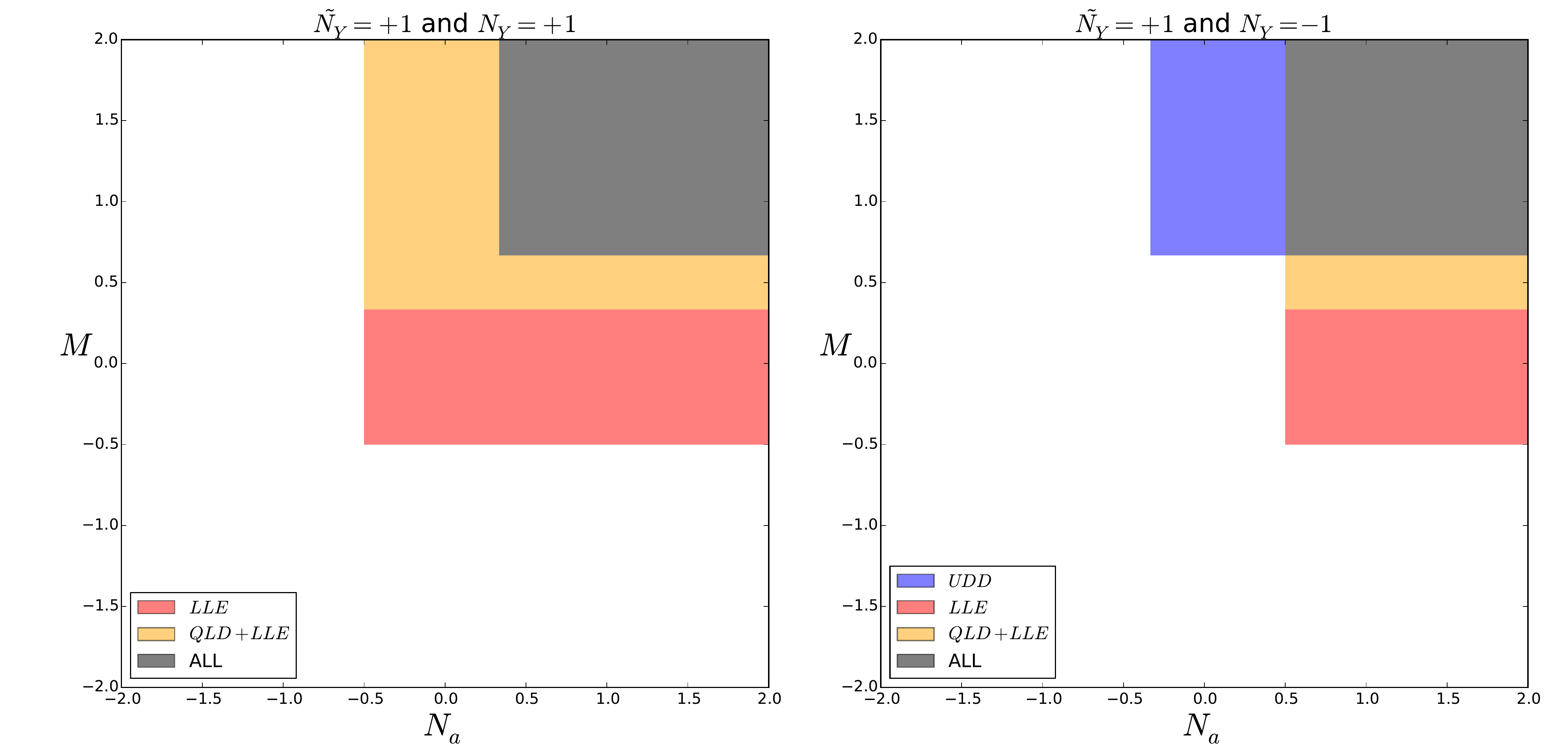}
\includegraphics[width=1.0\linewidth]{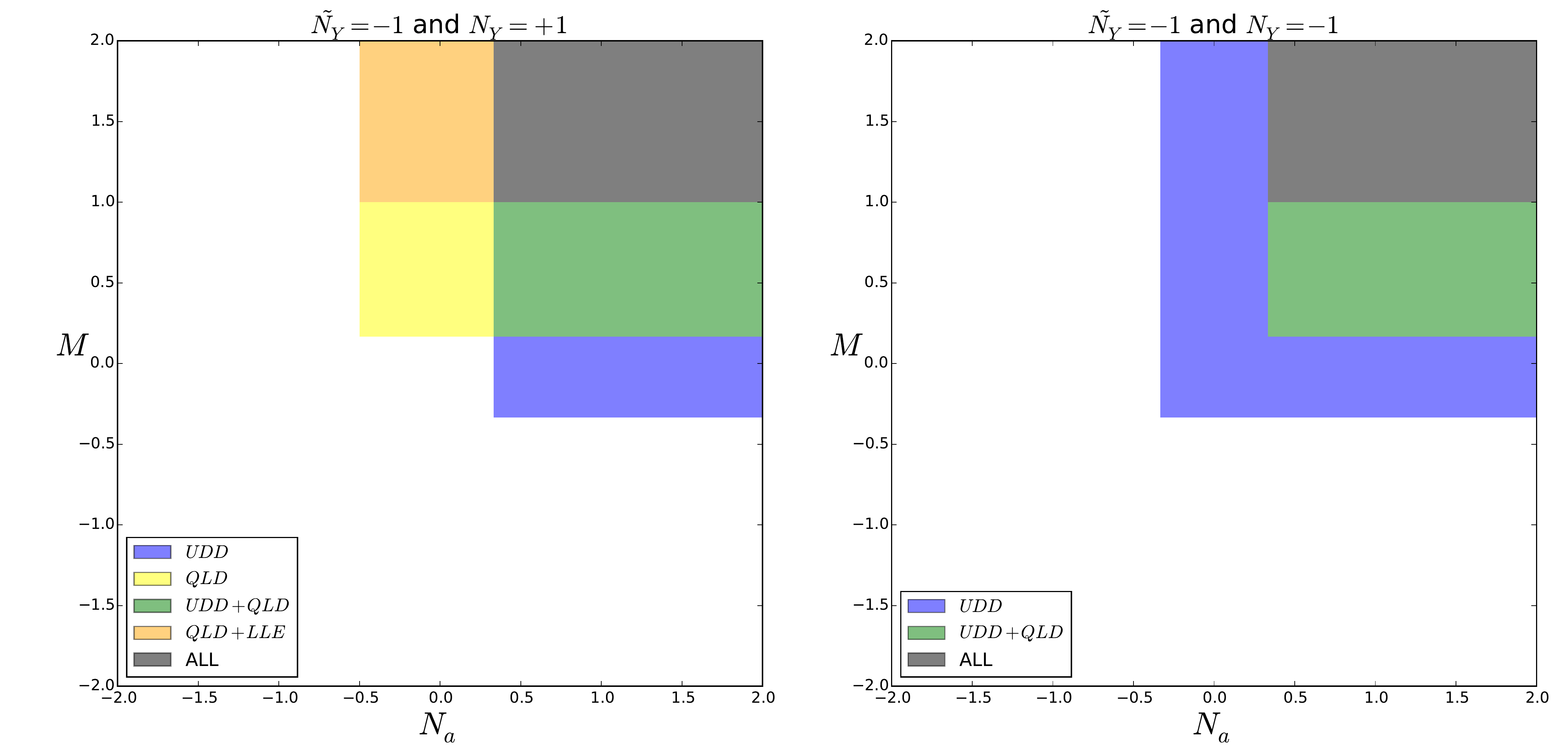}
\caption{Allowed regions in the parameter space for different RPV couplings. These figures should be seen in conjunction
with the operators presented in Table~\ref{table2}.}
\label{fig:regions_block}
\end{figure}

The local spectrum is assessed by local chiral index theorems \cite{Palti:2012aa}. In Appendix~\ref{app:LocalChirality} we outline the results for the constraints on flux densities such that different RPV points are allowed at a given $SO(12)$ enhancement point. These results are graphically presented in Figure \ref{fig:regions_block} and may be compared to the operators presented in Table~\ref{table2} in the semi-local approach. Thus, the green coloured region
is associated with  the $10_3\bar 5_1\bar 5_1$ operator of this Table, the blue colour with $10_1\bar 5_3\bar 5_3$, the pink  with 
$10_2\bar 5_4\bar 5_4$ and so on. 
Thus different regions of the parameter space can support different types of RPV interactions at a given enhancement point. We can then infer that in F-theory the allowed RPV interactions can, in principle, be only a subset of all possible RPV interactions.

In the limiting cases where only one coupling is turned on, one can derive bounds on its magnitude at the GUT scale from low-energy processes \cite{Allanach:1999ic}. In order to do so, one finds the bounds at the weak scale in the mass basis, performs a rotation to the weak basis and then evaluates the couplings at the GUT scale with the RGE. Since the effects of the rotation to the weak basis in the RPV couplings requires a full knowledge of the Yukawa matrices, we assume that the mixing only happens in the down-quark sector as we are not making any considerations regarding the up-quark sector in this work.  Table \ref{tab:AllRPVbounds} shows the upper bounds for the trilinear RPV couplings at the GUT scale.

The bounds presented in Table \ref{tab:AllRPVbounds} have to be understood as being derived under certain assumptions on mixing and points of the parameter space \cite{Barbier:2004ez, Allanach:2003eb}. For example, the bound on $\lambda_{12k}$ can be shown to have an explicit dependence on
\begin{equation}
	\frac{\tilde{m}_{e_{k,R}}}{100\mbox{ GeV}}
\end{equation}
where $\tilde{m}_{e_{k,R}}$ refers to a `right-handed' selectron soft-mass. The values presented in Table \ref{tab:AllRPVbounds}, as found in \cite{Allanach:1999ic}, were obtained by setting the soft-masses to $100$ GeV, which are ruled out by more recent LHC results \cite{Chun:2014jha,Deppisch:2012nb,Dreiner:2010ye,Khachatryan:2016iqn, Aad:2015iea, Aad:2015pfx} . By assuming heavier scalars, for example around $1$ TeV, we would then get the bounds in Table \ref{tab:AllRPVbounds} to be relaxed by one order of magnitude.

\begin{table}
	\begin{center}
	\begin{tabular}{c|c|c|c}
		$ijk$ & $\lambda_{ijk}$ & $\lambda'_{ijk}$ & $\lambda''_{ijk}$  \\ \hline
		111  &         -         & $1.5\times10^{-4}$ &          -           \\
		112  &         -         & $6.7\times10^{-4}$ & $4.1\times 10^{-10}$ \\
		113  &         -         &     $0.0059 $      &  $1.1\times10^{-8}$  \\
		121  &     $0.032 $      &      $0.0015$      & $4.1\times10^{-10}$  \\
		122  &     $0.032 $      &      $0.0015$      &          -           \\
		123  &     $0.032 $      &      $0.012$       & $1.3 \times 10^{-7}$ \\
		131  &     $0.041 $      &      $0.0027$      & $1.1\times 10^{-8}$  \\
		132  &     $0.041 $      &      $0.0027$      & $1.3\times 10^{-7}$  \\
		133  &     $0.0039$      & $4.4\times10^{-4}$ &          -           \\
		211  &      $0.032$      &      $0.0015$      &          -           \\
		212  &      $0.032$      &      $0.0015$      &       $(1.23)$       \\
		213  &      $0.032$      &      $0.016$       &       $(1.23)$       \\
		221  &         -         &      $0.0015$      &       $(1.23)$       \\
		222  &         -         &      $0.0015$      &          -           \\
		223  &         -         &      $0.049$       &       $(1.23)$       \\
		231  &      $0.046$      &      $0.0027$      &       $(1.23)$       \\
		232  &      $0.046$      &      $0.0028$      &       $(1.23)$       \\
		233  &      $0.046$      &      $0.048$       &          -           \\
		311  &      $0.041$      &      $0.0015$      &          -           \\
		312  &      $0.041$      &      $0.0015$      &       $0.099$        \\
		313  &     $0.0039$      &      $0.0031$      &       $0.015$        \\
		321  &     $0.046 $      &      $0.0015$      &       $0.099$        \\
		322  &      $0.046$      &      $0.0015$      &          -           \\
		323  &      $0.046$      &      $0.049$       &       $0.015$        \\
		331  &         -         &      $0.0027$      &       $0.015$        \\
		332  &         -         &      $0.0028$      &       $0.015$        \\
		333  &         -         &      $0.091$       &          -
	\end{tabular}
	\caption{
		Upper bounds of RPV couplings ($ijk$ refer to flavour/weak basis) at the GUT scale under the assumptions: 1) Only mixing in the down-sector, none in the Leptons; 2) Scalar masses 	$\tilde{m}=100$ GeV; 3) $\tan\beta(M_Z)=5$; and 4) Values in parenthesis refer to non-perturbative bounds, when these are stronger than the perturbative ones. This Table is reproduced from \cite{Allanach:1999ic}.\label{tab:AllRPVbounds}}
	\end{center}
\end{table}

\begin{figure}[h!]
\centering
\includegraphics[width=\linewidth]{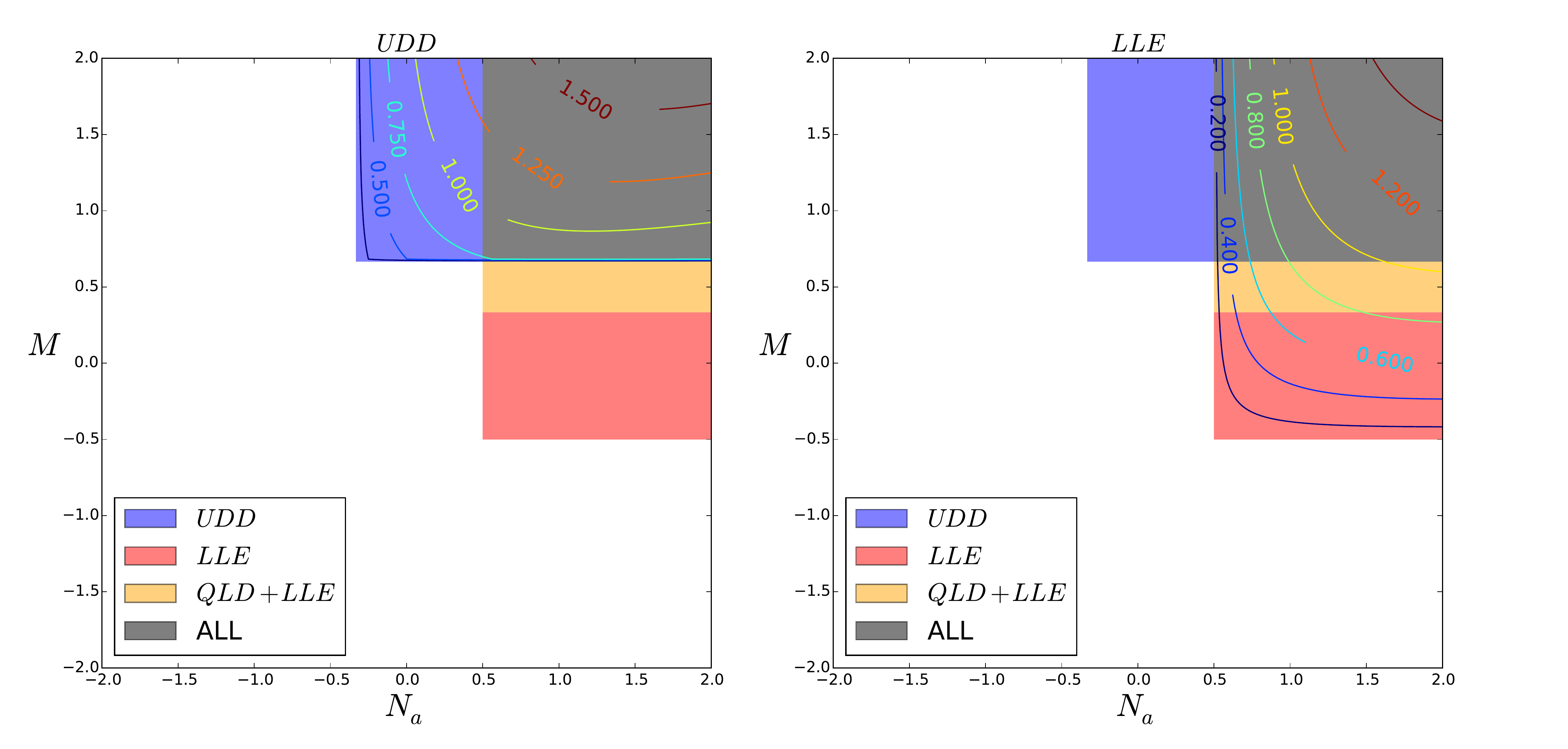}
\caption{Allowed regions in the parameter space for different RPV couplings with $\tilde{N_{Y}}=-N_Y=1$. We have also include the corresponding contours for the $u^cd^cd^c$ operator (left) and $LLe^c$ (right). }
\label{fig:regions_pm}
\end{figure}

\begin{figure}[h!]
\centering
\includegraphics[width=\linewidth]{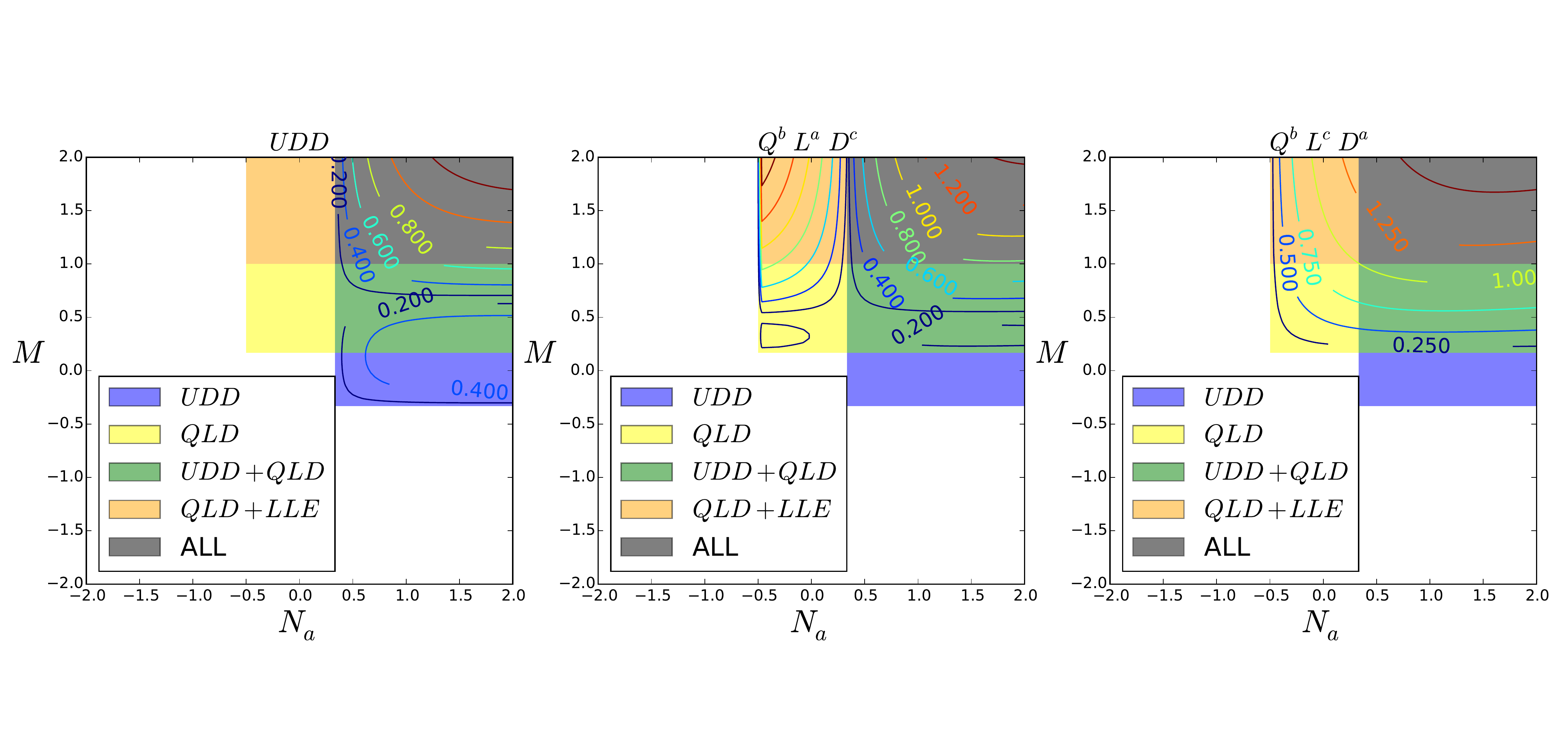}
\caption{Allowed regions in the parameter space for different RPV couplings with $N_{Y}=-\tilde{N_Y}=1$. We have also include the corresponding contours for the $u^cd^cd^c$ operator (left) and  $QLd^c$ (middle and right). The scripts a, b and c refer to which sector each state lives.}
\label{fig:regions_mp}
\end{figure}

\begin{figure}[h!]
	\centering
	\subfigure[$LLe^c$ regions with $\tilde{N_Y}=N_{Y}=1$\label{fig:regions_pp}]
	{\includegraphics[width=0.49\linewidth]{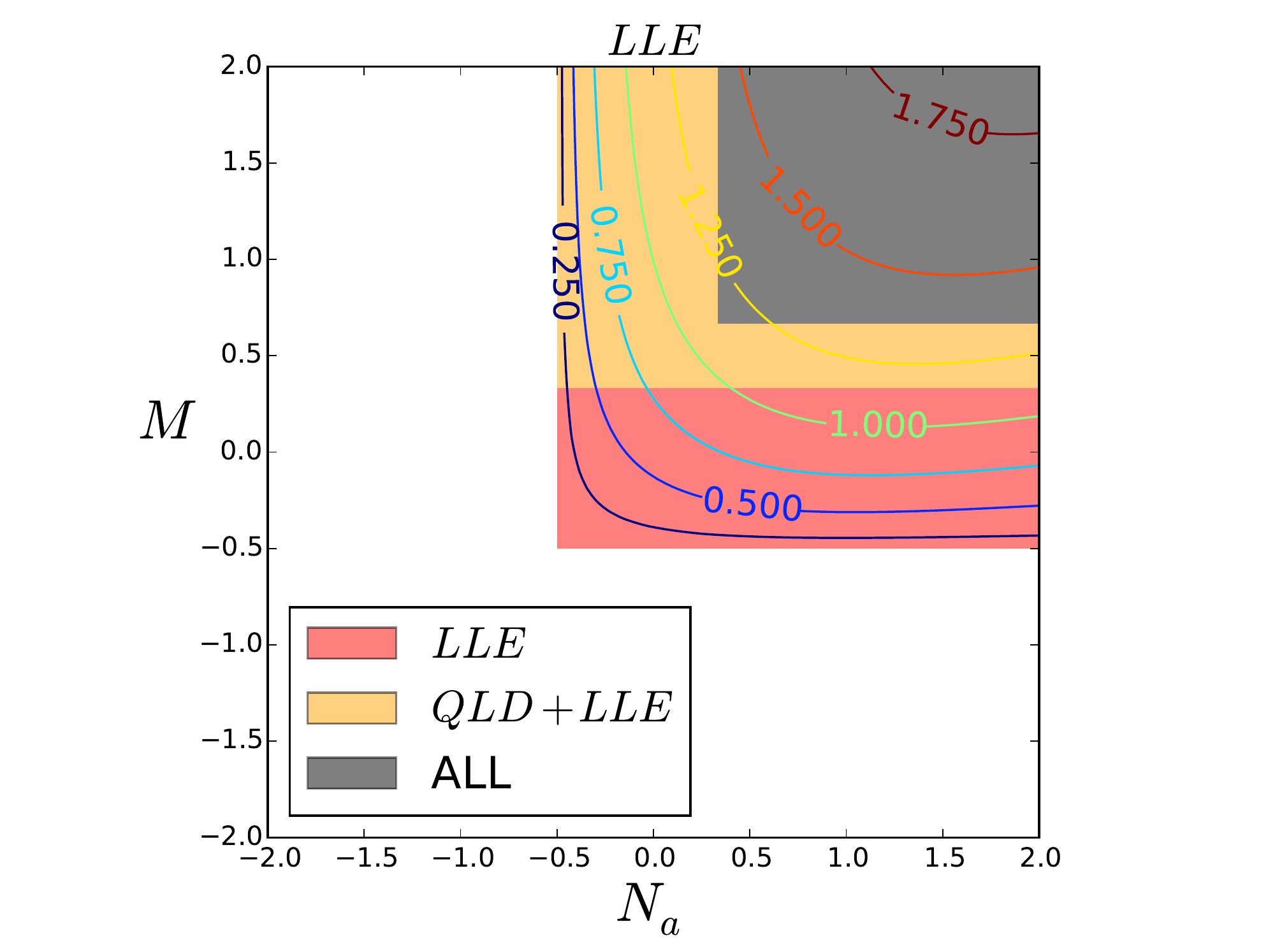}}
	\subfigure[$u^cd^cd^c$ regions with $\tilde{N_Y}=N_{Y}=-1$\label{fig:regions_mm}]
	{\includegraphics[width=0.49\linewidth]{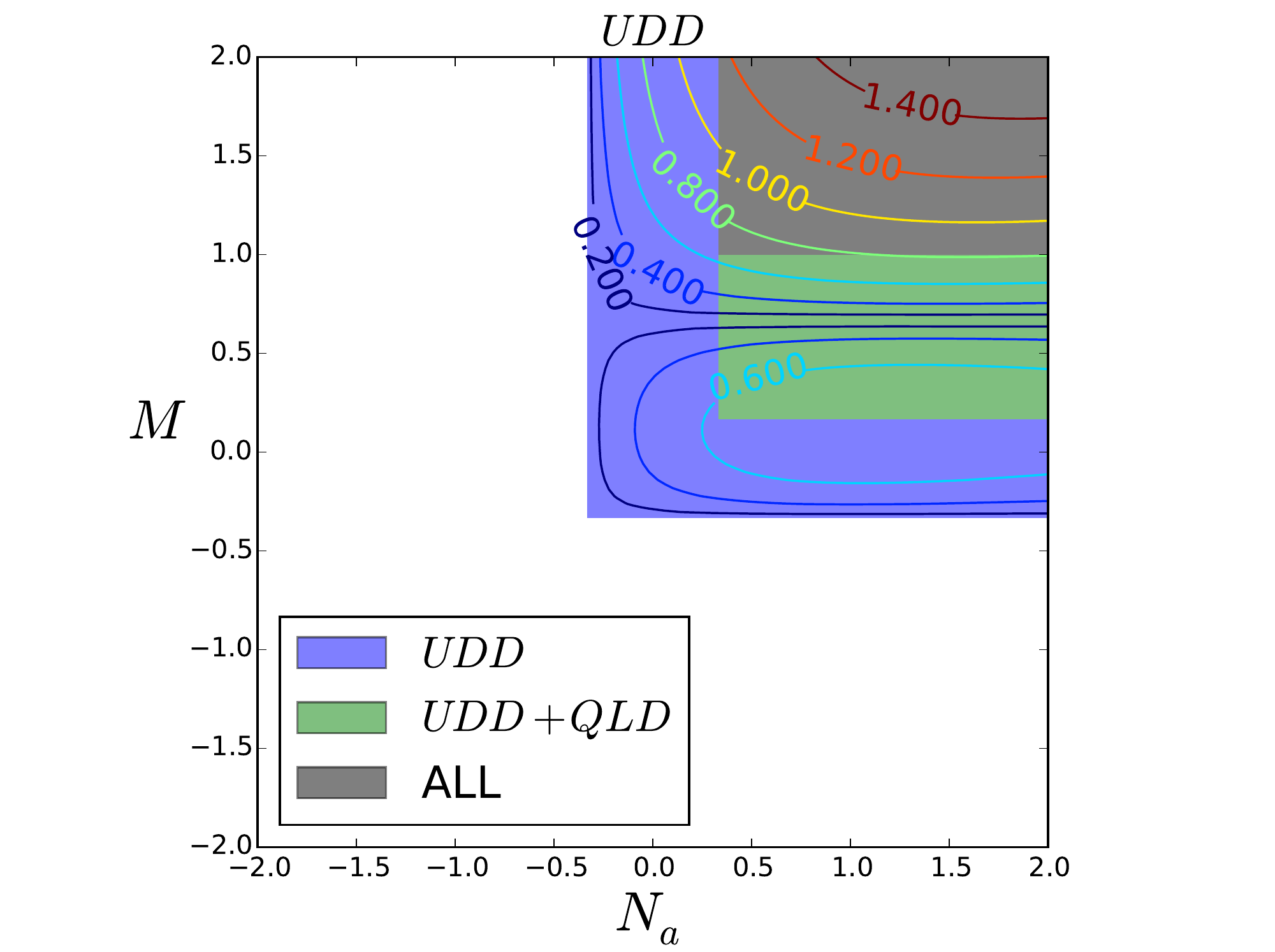}}
	\caption{Allowed regions in the parameter space for different RPV couplings.}
\end{figure}

The results show that the $\lambda$ type of coupling, corresponding to the $LLe^c$ interactions, is bounded to be $<0.05$ regardless of the indices taken. The red regions of Figures \ref{fig:regions_pp} and \ref{fig:regions_pm} show the magnitude of the coupling where it is allowed.
A similar analysis can be carried out for the remaining couplings. The $\lambda^\prime$ coupling, which measures the strength of the $LQd^c$ type of interactions,  can be seen in the yellow regions of Figure \ref{fig:regions_mp}. 
Finally, the derived values for $\lambda^{\prime\prime}$ coupling, related to the $u^cd^cd^c$ type of interactions, are shown in the blue regions of Figures \ref{fig:regions_mp} and \ref{fig:regions_mm}.
However these couplings shown are all expressed in units of $2 g_s^{1/2}\sigma$,
and so cannot yet be directly compared to the experimental limits.

In order to make contact with experiment we must eliminate the $2 g_s^{1/2}\sigma$ coefficient.
We do this by taking ratios of the couplings computed in this framework where the $2 g_s^{1/2}\sigma$ coefficient cancels in the ratio. The ratio between any RPV coupling and the bottom Yukawa at the GUT scale is given by
\begin{equation}
	r=\frac{y_{_{RPV}}}{y_{_b}} = \frac{y^\prime_{_{RPV}}}{y^\prime_b},
\end{equation}
as defined in \eref{eq:ybprime} and \eref{eq:yRPVprime}. This ratio can be used to assess the absolute strength of the RPV at the GUT scale as follows.

\begin{figure}[h!]
		\centering
		\subfigure[$LLe^c$ region with $N_Y=10$, $\tilde N_Y=0.1$\label{fig:figure20}]
		{\includegraphics[width=0.45\linewidth]{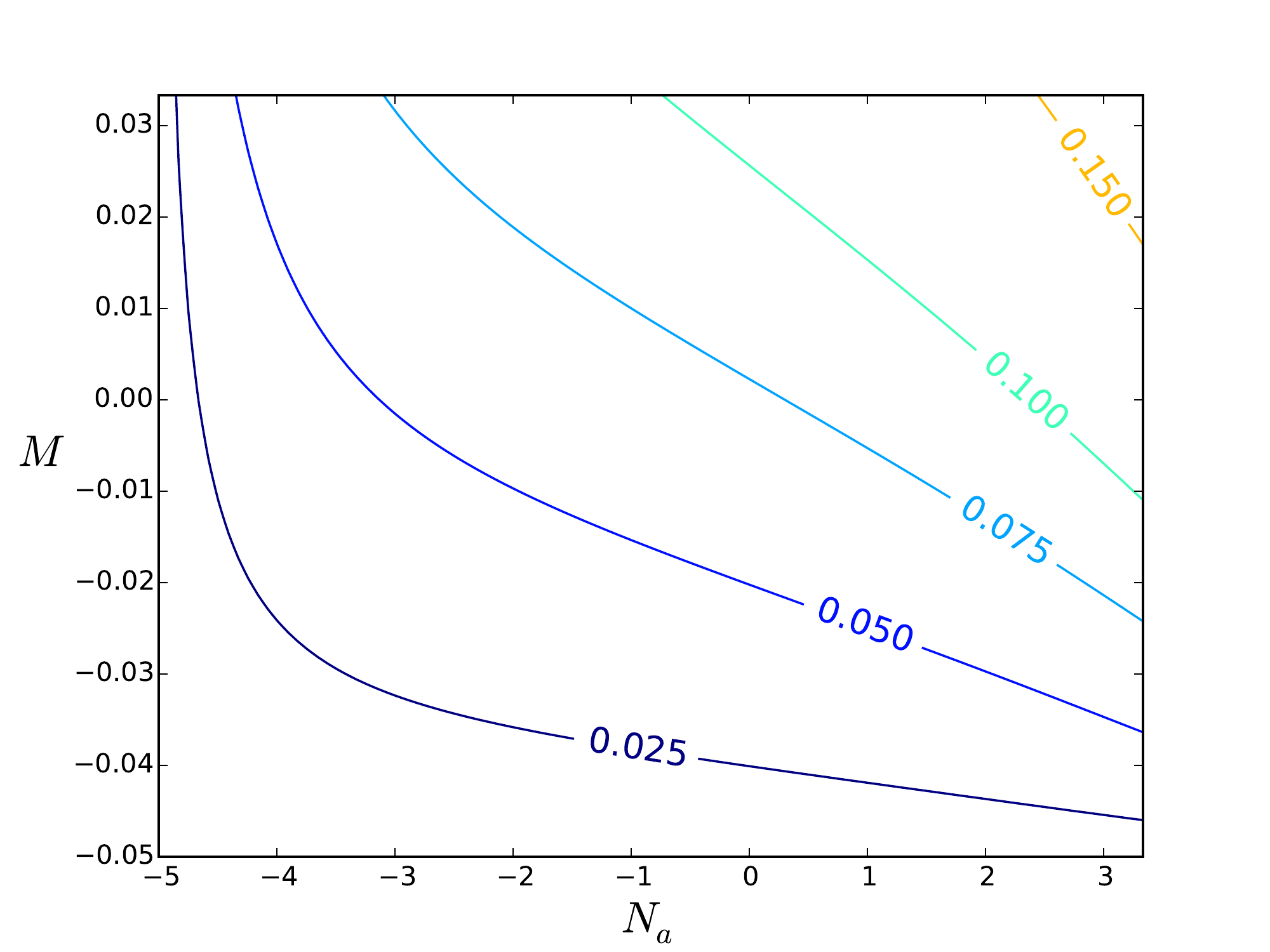}}
		\subfigure[$LLe^c$ region with $N_Y=-10$, $\tilde N_Y=0.1$\label{fig:figure21}]
		{\includegraphics[width=0.45\linewidth]{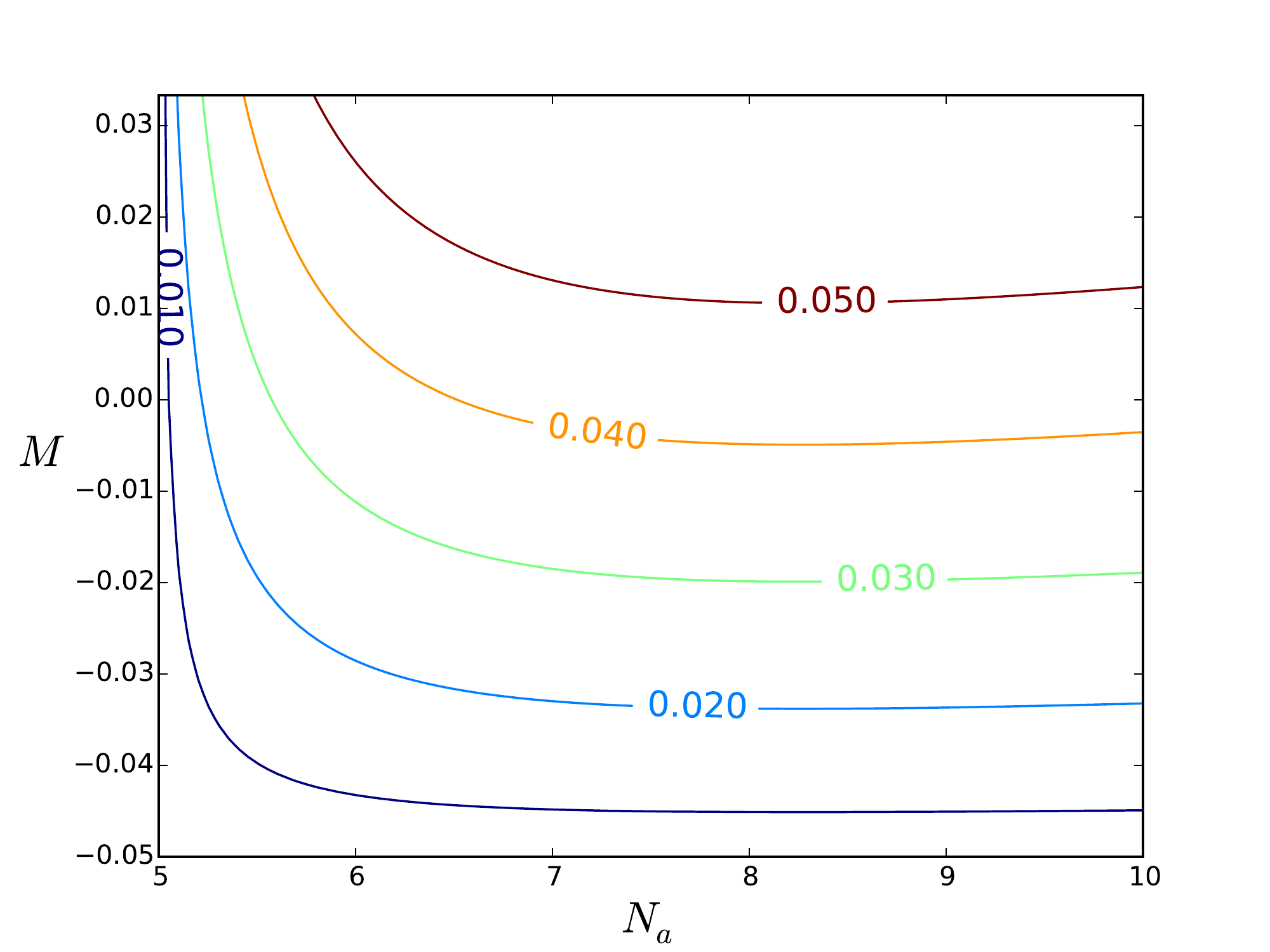}}
	 \vskip\baselineskip
	 		\subfigure[$QLd^c$ region with $N_Y=0.1$, $\tilde N_Y=-10$\label{fig:figure22}]
	 		{\includegraphics[width=0.45\linewidth]{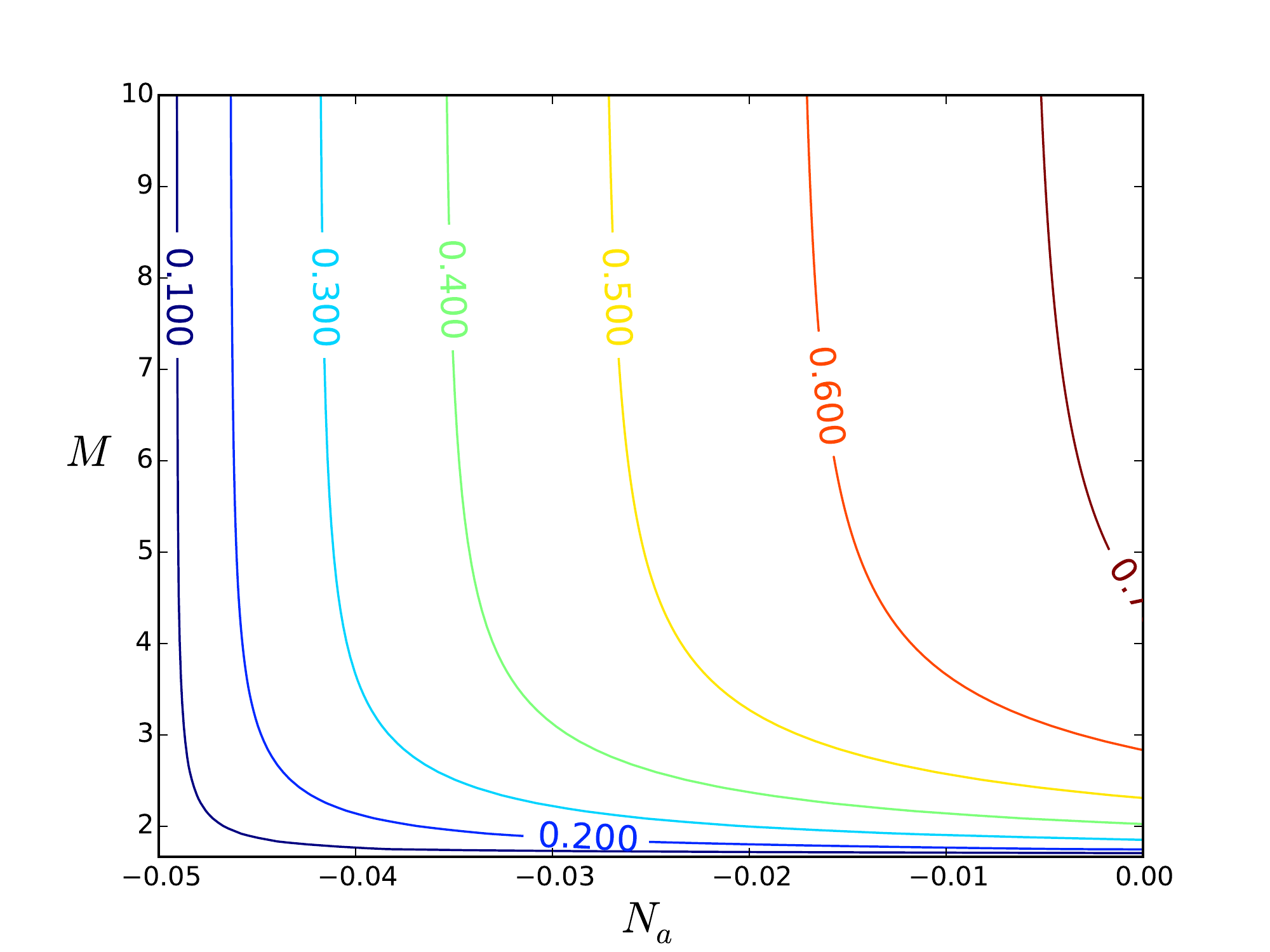}}
	 		\subfigure[$u^cd^cd^c$ region with $N_Y=-0.1$, $\tilde N_Y=-10$\label{fig:figure23}]
	 		{\includegraphics[width=0.45\linewidth]{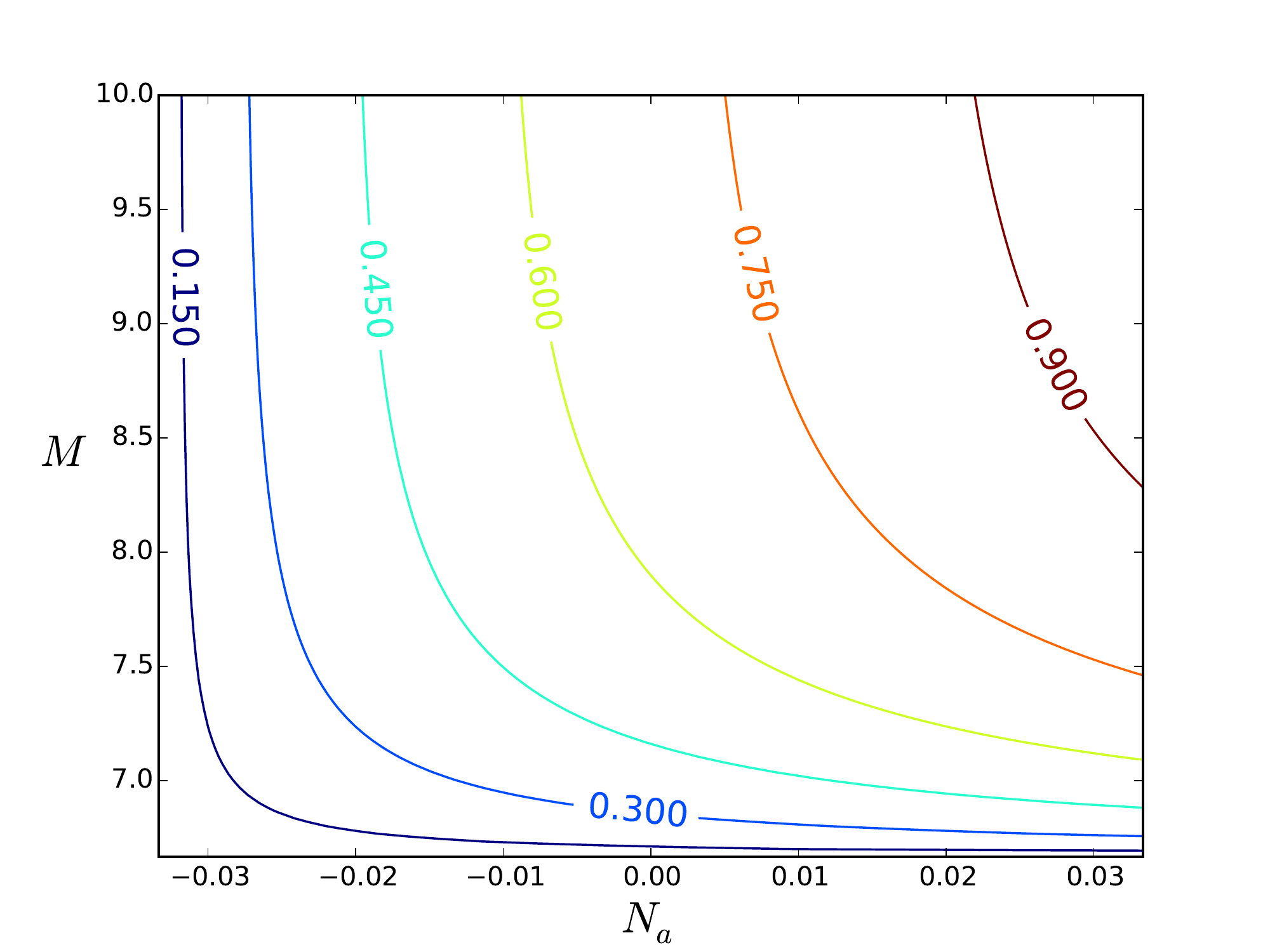}}
		\caption{$y_{_{RPV}}/y_{_b}$ ratio. The bottom Yukawa was computed in a parameter space point that returns a reasonable $y_b/y_\tau$ ratio  \cite{Font:2012wq}}
\end{figure}

First we assume that the RPV interaction is localised in an $SO(12)$ point far away from the bottom Yukawa point. This allows us to use different and independent flux densities at each point. We can then compute $y^\prime_b$ at a point in the parameter space where the ratio $y_b/y_\tau$ takes reasonable values, following \cite{Font:2012wq}. Finally we take the ratio, $r$.
In certain regions of the parameter space, $r$ is naturally smaller than $1$. This suppression of the RPV coupling in respect to the bottom Yukawa is shown in Figures \ref{fig:figure20}, \ref{fig:figure21}, \ref{fig:figure22}, and \ref{fig:figure23}, for different regions of the parameter space that allows for distinct types of RPV interactions.

Since $r$ is the ratio of both primed and unprimed couplings, respectively unphysical and physical, at the GUT scale, we can extend the above analysis to find the values of the physical RPV couplings at the GUT scale. To do so, we use low-energy, experimental, data to set the value of the bottom Yukawa at the weak scale for a certain value of $\tan \beta$. Next, we follow the study in \cite{Ross:2007az} to assess the value of the bottom Yukawa at the GUT scale through RGE runnings.

 \begin{figure}[h!]
	\centering
	\subfigure[$\lambda LLe^c$ region with $N_Y=10$, $\tilde N_Y=0.1$\label{fig:figure24}]
 	{\includegraphics[width=0.45\linewidth]{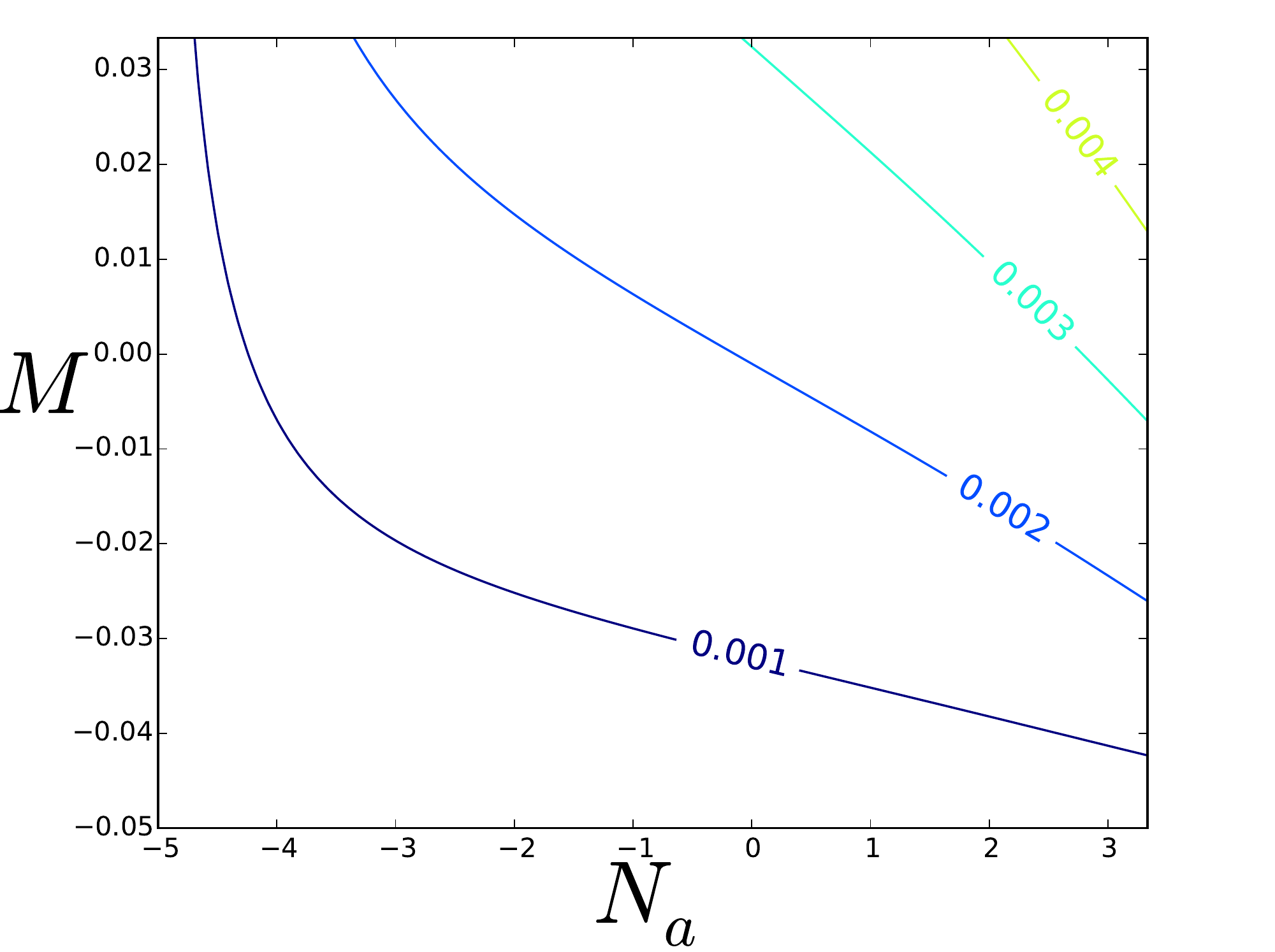}}
	\subfigure[$\lambda LLe^c$ region with $N_Y=-10$, $\tilde N_Y=0.1$\label{fig:figure25}]
	{\includegraphics[width=0.45\linewidth]{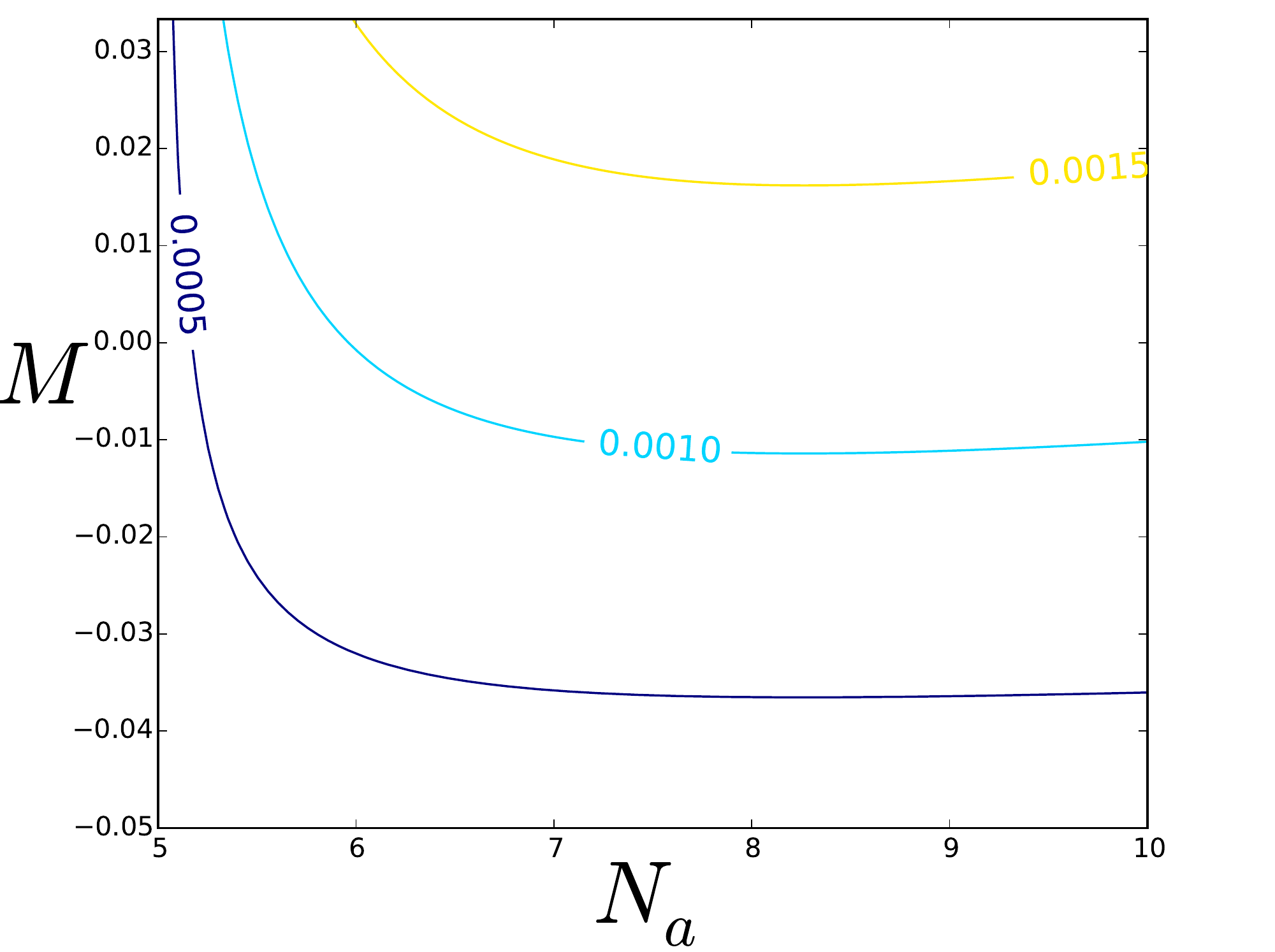}}
	\vskip\baselineskip
 	\subfigure[$\lambda' QLd^c$ region with $N_Y=0.1$, $\tilde N_Y=-10$\label{fig:figure26}]
	{\includegraphics[width=0.45\linewidth]{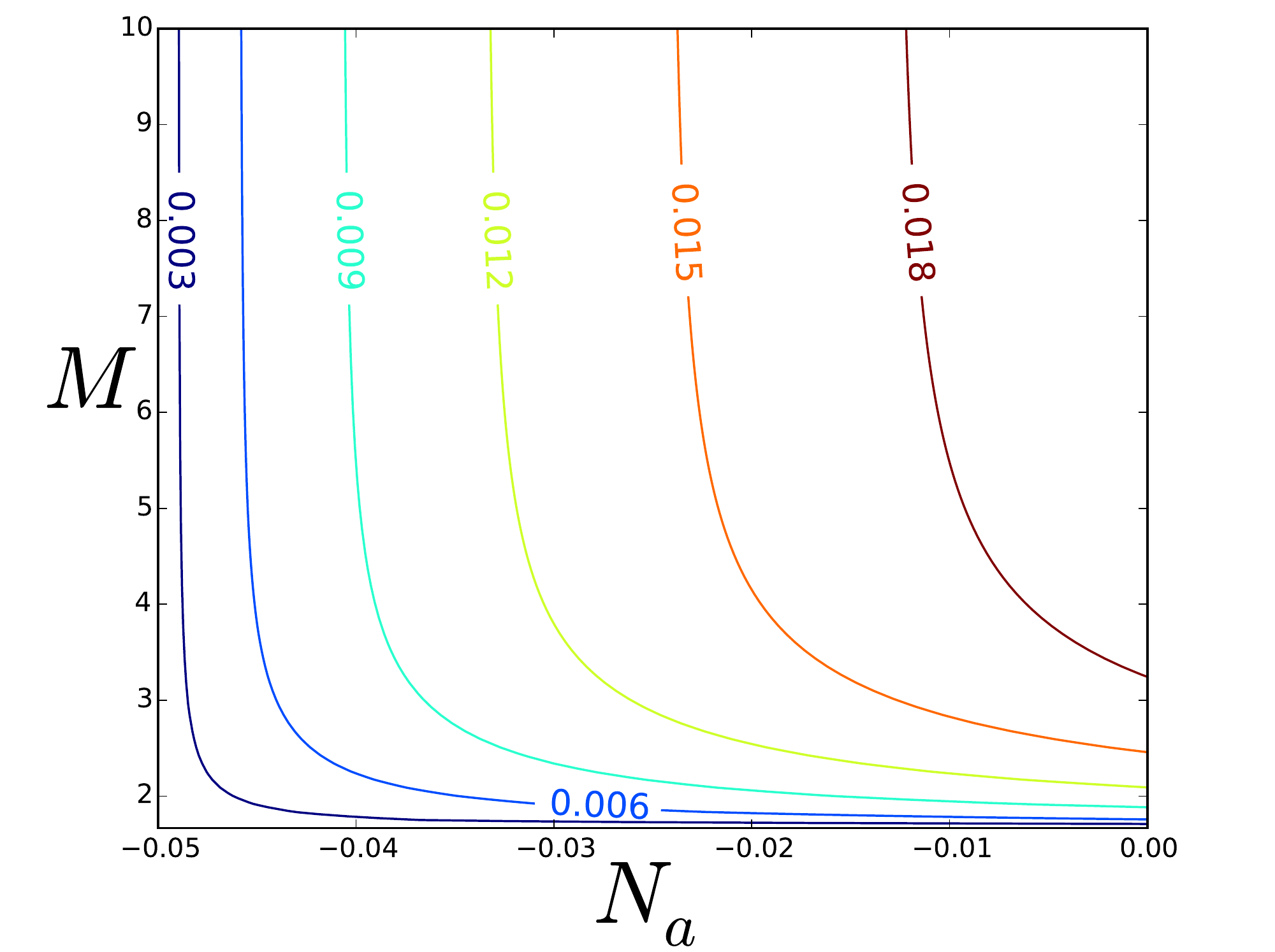}}
 	\subfigure[$\lambda ''u^cd^cd^c$ region with $N_Y=-0.1$, $\tilde N_Y=-10$\label{fig:figure27}]
 	{\includegraphics[width=0.45\linewidth]{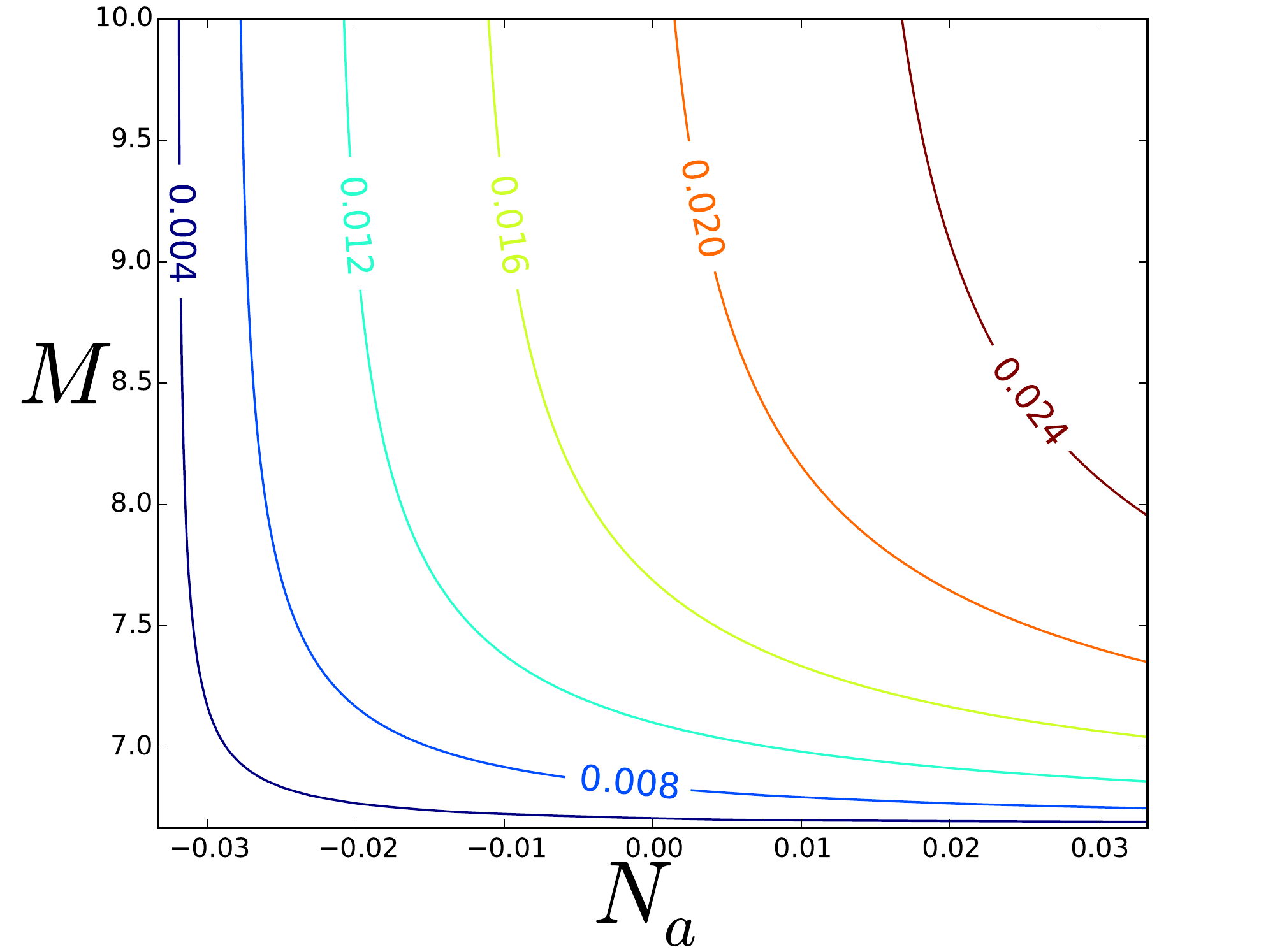}}
 	\caption{$y_{_{RPV}}$ at GUT scale for $\tan \beta=5$. The values here can be compared directly to the bounds presented in Table \ref{tab:AllRPVbounds}.}
 \end{figure}

In order to make a connection with the bounds in Table \ref{tab:AllRPVbounds}, we pick $\tan \beta = 5$ and we find $y_b (M_{_{GUT}}) \simeq 0.03$. The results for the value of the RPV couplings in different regions in the parameter space at the GUT scale are presented in Figures \ref{fig:figure24}, \ref{fig:figure25}, \ref{fig:figure26}, and \ref{fig:figure27}. These results show that, for any set of flavour indices, the strength of the coupling $\lambda $ related to an $LLe^c$ interaction is within the bounds. 
This means that this purely leptonic RPV operator, which violates lepton number 
but not baryon number, 
may be present with a sufficiently suppressed Yukawa coupling, according to our calculations. 
Therefore in the future lepton number violating processes could be observed.

By contrast, only for a subset of possible flavour index assignments for 
baryon number violating (but lepton number conserving) $u^cd^cd^c$ couplings 
are within the bounds in Table~\ref{tab:AllRPVbounds}.
The constraint on the first family up quark coupling $\lambda^{\prime\prime}_{1jk}$
for the $u^c_1d^c_jd^c_k$ interaction is so stringent, that this operator must only be permitted
for the cases $u^c_2d^c_jd^c_k$ and $u^c_3d^c_jd^c_k$ (corresponding to 
the two heavy up-type quarks $c^c, t^c$), assuming no up-type quark mixing.
However, if up-type quark mixing is allowed, then such operators could lead to 
an effective $u^c_1d^c_jd^c_k$ operator suppressed by small mixing angles,
in which case it could induce $n-\bar n$ oscillations~\cite{Karozas:2015zza}. 

Finally the $LQd^c$ operator with Yukawa coupling $\lambda'$ apparently must be avoided,
since according to our calculations, the value of $\lambda'$ that we predict exceeds the experimental 
limit by about an order of magnitude for all flavour indices, apart from $\lambda'_{333}$ coupling 
corresponding to the $L_3Q_3d^c_3$ operator. This implies that we should probably eliminate 
such operators which violate both baryon number and lepton number, using the flux mechanism
that we have described. However in some parts of parameter space, for certain 
flavour indices, such operators may be allowed leading to lepton number violating 
processes such as $K^+\rightarrow \pi^-e^+e^+$
and $D^+\rightarrow K^-e^+e^+$.

\section{Conclusions}
\label{conclusions}

In this paper we have provided the first dedicated study of R-parity violation (RPV) in F-theory semi-local
and local 
constructions based on the  $SU(5)$ grand unified theory (GUT) contained in the maximal subgroup 
$SU(5)_{GUT}\times SU(5)_{\bot} $ of an $E_8$ singularity associated with the elliptic fibration. 
Within this framework, we have tried to be 
as general  as possible, with the primary aim of making a bridge between F-theory and experiment.

We have focussed on semi-local and local F-theory $SU(5)$ constructions, where a non-trivial hypercharge flux breaks the GUT symmetry 
down to the Standard Model and  in addition  renders several GUT multiplets incomplete.  Acting on the Higgs 
curves this  novel mechanism can be regarded as the surrogate for the doublet-triplet  splitting of conventional GUTs. 
However, from a general perspective,  at the same time the hyperflux  may work  as a displacement mechanism,  removing 
certain components of GUT multiplets while accommodating fermion generations on other  matter curves.

 In the first part of the paper we considered semi-local constructions, focussing on
 F-theory $SU(5)_{GUT}$ models  which  are  classified  according  to  
 the discrete symmetries -- acting as identifications on the $SU(5)_{\bot}$ representations -- 
 and appearing as a subgroup of the maximal $SU(5)_{\bot}$ Weyl group $S_5$. Furthermore, we considered  phenomenologically
 appealing scenarios with the three fermion generations distributed on different matter curves
 and showed that RPV couplings  are a generic feature of such models. Upon introducing the 
 flux breaking mechanism, we classified all possible cases of incomplete GUT multiplets and examined the
 implications of their associated RPV couplings. Then we focused on the induced MSSM plus RPV Yukawa sector which involves only part of the MSSM allowed RPV operators as a consequence of the 
 missing components of the multiplets  projected out by the flux.  Next, we  tabulated all distinct cases and the type of physical process 
  (RPV or proton decay) that can arise from particular operators involving different types of incomplete multiplets.
  
  In the second part of the paper we computed the strength of the RPV Yukawa couplings, 
which mainly depend on  the topological properties of the internal space and are more or less independent of 
many details of a particular model, enabling us to work in a generic local F-theory setting.
Due to their physical relevance, we paid special attention to those couplings originating from  the $SU(5)$ 
operator $10\cdot \bar 5\cdot \bar 5$  in the presence of  general fluxes, which is realised at an $SO(12)$ point of enhancement.
Then, we applied the already developed F-theory techniques to calculate the numerical values of Yukawa couplings
for bottom, tau and RPV operators. Taking into account flux restrictions, which limit the types
of RPV operators that may appear simultaneously, we then  calculated ratios of Yukawa couplings, from 
which the physical RPV couplings at the GUT scale  can be determined.  We have explored the possible ranges of
the Yukawa coupling strengths of the $10\cdot \bar 5\cdot \bar 5$-type operators in a five-dimensional parameter space,  
corresponding to the number of the distinct flux parameters/densities associated with this superpotential term. 
Varying these  densities over a reasonable range of values, we have observed the tendencies of the
various Yukawa strengths with respect to the flux parameters and, to eliminate uncertainties from overall normalization constants,
we have computed the ratios of the RPV couplings to the bottom Yukawa one.
This way, using the experimentally determined mass of the bottom quark, we compared our results to limits 
on these couplings from experiment.

The results of this paper show firstly that, in semi-local F-theory constructions based on $SU(5)$ GUTs,
RPV is a generic feature, but may occur without proton decay, due to flux effects. Secondly, our calculations 
based on local F-theory constructions show that 
the value of the RPV Yukawa couplings at the GUT scale may be naturally 
suppressed over large regions of parameter space. 
Furthermore, we found that the existence of $LLe^c$ type of RPV interactions from F-Theory are expected to be within the current bounds. This implies that such lepton number violating operators 
could be present in the effective theory, but simply below current experimental limits, 
and so lepton number violation could be observed in the future.
Similarly, the baryon number violating operators $c^cd^c_jd^c_k$ 
and $t^cd^c_jd^c_k$ could also be present, leading to $n-\bar n$ oscillations.
Finally some $QLd^c$ operators could be present leading to 
lepton number violating 
processes such as $K^+\rightarrow \pi^-e^+e^+$
and $D^+\rightarrow K^-e^+e^+$.
In conclusion, our results suggest that RPV SUSY consistent with proton decay and current limits may be discovered in the future,
shedding light on the nature of F-theory constructions.

\vspace{1cm}
{\bf Acknowledgements}. SFK and GKL  are grateful to the Mainz Institute for Theoretical Physics (MITP)
 for its hospitality and its partial support during the completion of this work. SFK, AKM, and MCR are grateful to the University of Ioannina for its hospitality and its partial support during the completion of this work.
SFK acknowledges support from the STFC Consolidated grant ST/L000296/1 and the
European Union Horizon 2020 research and innovation programme under the Marie 
Sklodowska-Curie grant agreements InvisiblesPlus RISE No. 690575 and 
Elusives ITN No. 674896.
AKM is supported by  STFC studentship 1238679. MCR acknowledges support from the FCT under
the grant SFRH/BD/84234/2012.

\appendix

\section{Semi-local F-theory constructions: R-Parity violating couplings for the various monodromies\label{app:RPV}}

In this Appendix we examine the semi-local F-theory models in detail in order to demonstrate that RPV couplings are generic or at least common. To this end we note that:
\begin{enumerate}
	\item
	We want models with matter being distributed on different curves. This setup we call multi-curve models, in contrast to the models presented section 4 of \cite{Dudas:2010zb} and usually considered in other papers that compute Yukawa couplings.
	\item
	The models defined in this framework ``choose'' the $H_u$ assignment for us, since a tree-level, renormalizable, perturbative top-Yukawa requires the existence of the coupling
	\begin{equation}
	\mathbf{10}_a \mathbf{10}_a \mathbf{5}_b
	\end{equation}
	such that the perpendicular charges cancel out. As such, all the models listed above will have a definite assignment for the curve supporting $H_u$, and we do not assign the remaining MSSM states to curves, i.e. all the remaining $\mathbf{5}$ curves will be called $\mathbf{\overline{5}}_a$, making clear that they are either supporting some $\mathbf{\overline5}_M$ or $H_d$. Furthermore, we will refer to the $\mathbf{10}$ curve containing the top quark as $\mathbf{10}_M$.
	\item The indication for existence of tree-level, renormalizable, perturbative RPV is given by the fact we can find two couplings of the form
	\begin{align}
	\mathbf{10}_a \mathbf{\overline 5}_b \mathbf{\overline 5}_c \\
	\mathbf{10}_d \mathbf{\overline 5}_e \mathbf{\overline 5}_f  
	\end{align}
	for $(b,c) \neq (e,f)$, and $a,d$ unconstrained. This happens as $H_d$ cannot be both supported in one of the $\mathbf{\overline 5}_b$, $\mathbf{\overline 5}_c$ and at the same in one of the $\mathbf{\overline 5}_e$, $\mathbf{\overline 5}_f$.
	\item We do not make any comment on flux data. The above criteria can be evaded by switching off the fluxes such that the RPV coupling (once the assignment of $H_d$ to a curve is realised) disappears.
\end{enumerate}

With this in mind we study the possible RPV realisations in multi-curve  models.

\subsection{$2+1+1+1$}

In this case the spectral cover polynomial splits into four factors, three linear terms and a quadratic one.
 Also, due to the quadratic factor we impose a $Z_2$ monodromy. The bestiary of matter curves and their 
 perpendicular charges ($t_{i}$) is given in the Table~6.

\begin{table}[h!]\centering
\label{Tab2111}
	\begin{tabular}{rccccccccccc} \hline
Curve\, : &  $5_{H_u}$  & $5_1$ & $5_2$& $5_3$ & $5_4$& $5_5$ & $5_6$ & $10_M$ & $10_2$ & $10_3$ & $10_4$ \\ 
Charge\, :&  ${\scriptstyle -2t_1}$  & ${\scriptstyle -t_{1}-t_3}$ & ${\scriptstyle -t_{1}-t_4}$& ${\scriptstyle -t_{1}-t_5}$ & ${\scriptstyle -t_{3}-t_4}$& ${\scriptstyle t_{3}-t_5}$ & ${\scriptstyle -t_{4}-t_5}$ & ${\scriptstyle t_1}$ & ${\scriptstyle t_3}$ & ${\scriptstyle t_4}$ & ${\scriptstyle t_5}$\\ \hline
	\end{tabular} \caption{Matter curves and the corresponding $U(1)$ charges for the case of a $2+1+1+1$ spectral cover split. Note that because of the $Z_2$ monodromy we have $t_{1}\longleftrightarrow{t_2}$.}
\end{table}

In this model RPV is expected to be generic as we have the following terms

\begin{equation}
\mathbf{10}_4\mathbf{\overline 5}_1\mathbf{\overline 5}_2  , \
\mathbf{10}_3\mathbf{\overline 5}_1\mathbf{\overline 5}_3  , \
\mathbf{10}_M\mathbf{\overline 5}_1\mathbf{\overline 5}_6  , \
\mathbf{10}_2\mathbf{\overline 5}_2\mathbf{\overline 5}_3  , \
\mathbf{10}_M\mathbf{\overline 5}_2\mathbf{\overline 5}_5  , \
\mathbf{10}_M \mathbf{\overline 5}_3\mathbf{\overline 5}_4 
\end{equation}

\subsection{$2+2+1$}

Here the spectral cover polynomial splits into three factors, it is the product of two quadratic terms and a linear one. We can impose a $Z_{2}\times{Z_{2}}$ monodromy which leads to the following identifications between the weights,($t_{1}\leftrightarrow{t_2}$) and ($t_{3}\leftrightarrow{t_4}$) .
In this case there are two possible assignments for $H_u$  (and $\mathbf{10}_M$), as we can see in Table~7.

\begin{table}[h!]\centering
	\begin{tabular}{rcccccccc}
	\hline \multicolumn{9}{c}{ case 1} \\
Curve  &  $5_{H_u}$  & $5_1$ & $5_2$& $5_3$ & $5_4$ & $10_M$ & $10_2$ & $10_3$  \\
Charge &  ${\scriptstyle -2t_1}$  & ${\scriptstyle -t_{1}-t_3}$ & ${\scriptstyle -t_{1}-t_{5}}$& ${\scriptstyle -t_{3}-t_{5}}$ & ${\scriptstyle -2t_{3}}$& ${\scriptstyle t_{1}}$ & ${\scriptstyle -t_{3}}$ & ${\scriptstyle t_5}$ \\ \hline
 \multicolumn{9}{c}{ case 2} \\
Curve  &  $5_{H_u}$  & $5_1$ & $5_2$& $5_3$ & $5_4$ & $10_M$ & $10_2$ & $10_3$  \\
Charge &  ${\scriptstyle -2t_3}$  & ${\scriptstyle -t_{1}-t_3}$ & ${\scriptstyle -t_{1}-t_5}$& ${\scriptstyle -t_{3}-t_5}$ & ${\scriptstyle -2t_{1}}$& ${\scriptstyle t_{3}}$ & ${\scriptstyle -t_{1}}$ & ${\scriptstyle t_5}$ \\\hline
	\end{tabular} \caption{The scenario of a $2+2+1$ spectral cover split with the corresponding matter curves and $U(1)$ charges. Note that we have two possible cases.}
\end{table}\label{Tab221}

\subsubsection{$2+2+1$ case 1}

The bestiary of matter curves and their perp charges is given in the upper half table of Table~7.

In this model RPV is expected to be generic as we have the following terms
\begin{equation}
\mathbf{10}_2\mathbf{\overline 5}_1\mathbf{\overline 5}_2  , \
 \mathbf{10}_M \mathbf{\overline 5}_1\mathbf{\overline 5}_3, \
\mathbf{10}_M\mathbf{\overline 5}_2\mathbf{\overline 5}_4   , \
\mathbf{10}_3 \mathbf{\overline 5}_1\mathbf{\overline 5}_1 
\end{equation}

Notice that if $\mathbf{\overline 5}_1$ contains only one state, then the last coupling is absent due to anti-symmetry of SU(5) contraction.

\subsubsection{$2+2+1$ case 2}

The bestiary of matter curves and their perp charges is given in the lower
half table of Table~7.

In this model RPV is expected to be generic as we have the following terms
\begin{equation}
\mathbf{10}_M\mathbf{\overline 5}_1\mathbf{\overline 5}_2  , \
\mathbf{10}_2\mathbf{\overline 5}_1\mathbf{\overline 5}_3  , \
\mathbf{10}_M \mathbf{\overline 5}_3\mathbf{\overline 5}_4 , \
\mathbf{10}_3\mathbf{\overline 5}_1\mathbf{\overline 5}_1  
\end{equation}

Notice that if $\mathbf{\overline 5}_1$ contains only one state, then the last coupling is absent due to anti-symmetry of SU(5) contraction.

\subsection{$3+1+1$}

In this scenario the splitting of the spectral cover leads to a cubic and two linear factors. We can impose a $Z_3$ monodromy for the roots of the cubic polynomial. The bestiary of matter curves and their perpendicular charges is given in Table~8:
\begin{table}[h!]\centering
	\begin{tabular}{rccccccc}\hline
Curve  &  $5_{H_u}$  & $5_1$ & $5_2$& $5_3$ & $10_M$ & $10_2$ & $10_3$  \\
Charge &  ${\scriptstyle -2t_1}$  & ${\scriptstyle -t_{1}-t_4}$ & ${\scriptstyle -t_{1}-t_5}$& ${\scriptstyle -t_{4}-t_5}$ & ${\scriptstyle t_1}$ & ${\scriptstyle t_4}$ & ${\scriptstyle t_5}$\\\hline
	\end{tabular} \caption{Matter curves and the corresponding $U(1)$ charges for the case of a $3+1+1$ spectral cover split. Note that we have impose a $Z_3$ monodromy.}
\end{table}\label{tab311}

In this model R-parity violation is not immediately generic as we only have
\begin{equation}
\mathbf{10}_2\mathbf{\overline 5}_1\mathbf{\overline 5}_2 , \
\mathbf{10}_M\mathbf{\overline 5}_1\mathbf{\overline 5}_3 
\end{equation}
and as such assigning $H_d$ to $\mathbf{\overline 5}_1$ avoids tree-level, renormalizable, perturbative RPV.

\subsection{$3+2$}

These type of models are in general very constrained because of the large monodromies which  leads to a low number of matter curves. 

In this case there are two possible assignments for $H_u$  (and $\mathbf{10}_M$), as described in  
Table~9.

\begin{table}[h!]\centering
	\begin{tabular}{rccccc}
	\hline \multicolumn{6}{c}{case 1} \\
Curve  &  $5_{H_u}$  & $5_2$ & $5_3$& $10_M$ & $10_2$ \\
Charge &  ${\scriptstyle -2t_1}$  & ${\scriptstyle -t_{1}-t_3}$ & ${\scriptstyle -2t_3}$& ${\scriptstyle t_1}$ & ${\scriptstyle t_{3}}$ \\ \hline
 \multicolumn{6}{c}{ case 2} \\ 
Curve  &  $5_{H_u}$  & $5_2$ & $5_3$& $10_M$ & $10_2$ \\
Charge &  ${\scriptstyle -2t_3}$  & ${\scriptstyle -t_{1}-t_3}$ & ${\scriptstyle -2t_1}$& ${\scriptstyle t_3}$ & ${\scriptstyle t_{1}}$ \\
\hline
	\end{tabular} \caption{The two possible cases in the scenario of a $3+2$ spectral cover split, the matter curves and the corresponding $U(1)$ charges.}
\end{table}\label{tab32}

\subsubsection{$3+2$ case 1}

The matter curves content is given in the upper half of Table~9 (case 1).

Possible RPV couplings are
\begin{equation}
\mathbf{10}_M \overline{\mathbf{5}}_2\overline{\mathbf{5}}_3\,,\, \mathbf{10}_2\overline{\mathbf{5}}_2\overline{\mathbf{5}}_2
\end{equation}

Notice that if $\mathbf{\overline 5}_2$ contains only one state, then the last coupling is absent due to anti-symmetry of SU(5) contraction.

\subsubsection{$3+2$ case 2}

This second scenario is referred as case 2 in the lower half of Table~9.

Only one coupling
\begin{equation}
\mathbf{10}_M\overline{\mathbf{5}}_2\overline{\mathbf{5}}_2
\end{equation}
which is either RPV or is absent.
Notice that if $\mathbf{\overline 5}_2$ contains only one state, then the last coupling is absent due to anti-symmetry of SU(5) contraction.

\section{Local F-theory constructions: local chirality constraints on flux data and R-Parity violating operators\label{app:LocalChirality}}

The chiral spectrum of a matter curve is locally sensitive to the flux data. This is happens as there is a notion of local chirality due to local index theorems \cite{Palti:2012aa,Font:2013ida}. The presence of a chiral state in a sector with root $\rho$ is given if the matrix
\[
m_\rho =\left(
\begin{array}{ccc}
-{q_P} & {q_S} & i {m}^2 {q_{z_1}} \\
{q_S} & {q_P} & i {m}^2 {q_{z_2}} \\
-i {m}^2 {q_{z_1}} & -i {m}^2 {q_{z_2}} &
0 \\
\end{array}
\right)
\]
with $q_i$ presented in Table \ref{tab:SO12curvesWflux}, has positive determinant
\begin{equation}\label{eq:detmrho}
\det m_\rho > 0 .
\end{equation}

As such, if we want a certain RPV coupling to be present, then the above condition has to be satisfied for the three states involved in the respective interaction at the $SO(12)$ enhancement point. For example, in order for the emergence of an $QLd^c$ type of RPV interaction, locally the spectrum has to support a $Q$, a $L$, and a $d^c$ states. The requirement that at a single point Equation \eqref{eq:detmrho} hold for each of these states imposes constraints on the values of the flux density parameters.

Therefore, while RPV effects in general include all three operators - $QLd^c$, $u^cd^cd^c$, $LLe^c$ - there are regions of the parameter space that allow for the elimination of some or all of the couplings. These are in principle divided into four regions, depending on the sign of the parameters $\tilde{N}_Y$ and $N_Y$. In the appendix we present the resulting regions of the parameter space and which operators are allowed in each.

\subsection{$\tilde{N}_Y\le0$}

For $\tilde N_Y \le0$, the conditions on the flux density parameters for which each RPV interaction is turned on are
\begin{align*}
QLd^c:\,\,\,&M>\frac{-\tilde{N}_Y}{6}\\
&N_a-N_b>\frac{-N_Y}{2}\\
u^c d^c d^c:\,\,\,&M>\frac{\tilde{N}_Y}{3}\\
&N_a-N_b>-\frac{N_Y}{3}\\
LLe^c:\,\,\,&M>-\tilde{N}_Y\\
&N_a-N_b>\frac{-N_Y}{2}
\end{align*}

Depending on the sign of $N_Y$, the above conditions define different regions of the flux density parameter space. These are presented in Tables~\ref{tab:localNYtildnegNYpos} and \ref{tab:localNYtildnegNYneg}.


\begin{table}[h!]\centering

\begin{tabular}{c|c|c|c|c}

$-$ & $M<\frac{\tilde{N}_Y}{3}$  & $\frac{\tilde{N}_Y}{3}<M<\frac{-\tilde{N}_Y}{6}$ & $\frac{-\tilde{N}_Y}{6}<M<-\tilde{N}_Y$ & $-\tilde{N}_Y<M$ \\\hline

$(N_a-N_b)<\frac{-N_Y}{2}$ &None &None &None & None\\\hline

$\frac{-N_Y}{2}<(N_a-N_b)<\frac{N_Y}{3}$ & None&None & $QLd^c$ &$QLd^c$, $LLe^c$ \\\hline

$\frac{N_Y}{3}<(N_a-N_b)$ & None & $u^cd^cd^c$ & $QLd^c$, $u^cd^cd^c$& All

\end{tabular}
\caption{Regions of the parameter space and the respective RPV operators supported for $\tilde{N}_Y\le0$, $N_Y >0$}\label{tab:localNYtildnegNYpos}
\end{table}


\begin{table}[h!]\centering

\begin{tabular}{c|c|c|c|c}

$-$ & $M<\frac{\tilde{N}_Y}{3}$  & $\frac{\tilde{N}_Y}{3}<M<\frac{-\tilde{N}_Y}{6}$ & $\frac{-\tilde{N}_Y}{6}<M<-\tilde{N}_Y$ & $-\tilde{N}_Y<M$ \\\hline

$(N_a-N_b)<\frac{N_Y}{3}$ &None &None &None & None\\\hline

$\frac{N_Y}{3}<(N_a-N_b)<\frac{-N_Y}{2}$ & None& $u^cd^cd^c$ & $u^cd^cd^c$ & $u^cd^cd^c$\\\hline

$\frac{-N_Y}{2}<(N_a-N_b)$& None & $u^cd^cd^c$ & $QLd^c$, $u^cd^cd^c$ & All

\end{tabular}
\caption{Regions of the parameter space and the respective RPV operators supported for $\tilde{N}_Y\le0$, $N_Y <0$}\label{tab:localNYtildnegNYneg}
\end{table}

\subsection{$\tilde{N}_Y>0$}

For $\tilde N_Y > 0$, the conditions on the flux density parameters for which each RPV interaction is turned on are
\begin{align*}
QLd^c:\,\,\,&M>\frac{\tilde{N}_Y}{3}\\
&N_a-N_b>\frac{-N_Y}{2}\\
u^cd^cd^c:\,\,\,&M>\frac{2\tilde{N}_Y}{3}\\
&N_a-N_b>-\frac{N_Y}{3}\\
LLe^c:\,\,\,&M>\frac{-\tilde{N}_Y}{2}\\
&N_a-N_b>\frac{-N_Y}{2}
\end{align*}

Depending on the sign of $N_Y$, the above conditions define different regions of the flux density parameter space. These are presented in Tables \ref{tab:localNYtildposNYpos} and \ref{tab:localNYtildposNYneg}.


\begin{table}[h!]\centering

\begin{tabular}{c|c|c|c|c}

$-$ & $M<-\frac{-\tilde{N}_Y}{2}$  & $-\frac{-\tilde{N}_Y}{2}<M<\frac{\tilde{N}_Y}{3}$ & $\frac{\tilde{N}_Y}{3}<M<\frac{2\tilde{N}_Y}{3}$ & $\frac{2\tilde{N}_Y}{3}<M$    \\\hline

$(N_a-N_b)<\frac{-N_Y}{2}$ &None &None &None & None\\\hline

$\frac{-N_Y}{2}<(N_a-N_b)<\frac{N_Y}{3}$ & None& $LLe^c$ & $QLd^c$, $LLe^c$ & $QLd^c$, $LLe^c$\\\hline

$\frac{N_Y}{3}<(N_a-N_b)$ & None & $LLe^c$ & $QLd^c$, $LLe^c$ & All

\end{tabular}
\caption{Regions of the parameter space and the respective RPV operators supported for $\tilde{N}_Y>0$, $N_Y >0$}\label{tab:localNYtildposNYpos}
\end{table}


\begin{table}[h!]\centering

\begin{tabular}{c|c|c|c|c}

$-$ & $M<-\frac{-\tilde{N}_Y}{2}$  & $-\frac{-\tilde{N}_Y}{2}<M<\frac{\tilde{N}_Y}{3}$ & $\frac{\tilde{N}_Y}{3}<M<\frac{2\tilde{N}_Y}{3}$ & $\frac{2\tilde{N}_Y}{3}<M$    \\\hline

$(N_a-N_b)<\frac{N_Y}{3}$ &None &None &None & None\\\hline

$\frac{N_Y}{3}<(N_a-N_b)<\frac{-N_Y}{2}$ & None& None & None & $u^cd^cd^c$ \\\hline

$\frac{-N_Y}{2}<(N_a-N_b)$& None & $LLe^c$ & $QLd^c$, $LLe^c$ & All

\end{tabular}
\caption{Regions of the parameter space and the respective RPV operators supported for $\tilde{N}_Y>0$, $N_Y <0$}\label{tab:localNYtildposNYneg}
\end{table}

 \end{document}